\documentclass{article}

\usepackage{PRIMEarxiv}

\usepackage[utf8]{inputenc} % allow utf-8 input
\usepackage[T1]{fontenc}    % use 8-bit T1 fonts
\usepackage{hyperref}       % hyperlinks
\usepackage{url}            % simple URL typesetting
\usepackage{booktabs}       % professional-quality tables
\usepackage{amsfonts}       % blackboard math symbols
\usepackage{nicefrac}       % compact symbols for 1/2, etc.
\usepackage{microtype}      % microtypography
\usepackage{lipsum}
\usepackage{fancyhdr}       % header
\usepackage{graphicx}       % graphics
\graphicspath{{media/}}     % organize your images and other figures under media/ folder

\usepackage{graphicx}%
\usepackage{multirow}%
\usepackage{amsmath,amssymb,amsfonts}%
\usepackage{amsthm}%
\usepackage{mathrsfs}%
\usepackage[title]{appendix}%
\usepackage{xcolor}%
\usepackage{textcomp}%
\usepackage{manyfoot}%
\usepackage{booktabs}%
\usepackage{algorithm}%
\usepackage{algorithmicx}%
\usepackage{algpseudocode}%
\usepackage{listings}%
\usepackage{natbib}
\usepackage{tabulary}
\newcolumntype{K}[1]{>{\centering\arraybackslash}p{#1}}

\usepackage[utf8]{inputenc}
\usepackage[T1]{fontenc}
\usepackage{mathtools}
\usepackage[thinc]{esdiff}

\usepackage{graphicx}
\usepackage{lipsum}
\usepackage[section]{placeins}

\usepackage{caption}

\newcommand{\citetmiddle}[1]{(\bgroup \citeauthor{#1} \citeyear{#1}\egroup)}

\usepackage{tabulary}
\newcolumntype{K}[1]{>{\centering\arraybackslash}p{#1}}

\usepackage[utf8]{inputenc}
\usepackage[T1]{fontenc}
\usepackage{mathtools}
\usepackage[thinc]{esdiff}

\usepackage{graphicx}
\usepackage{lipsum}
\usepackage[section]{placeins}

\usepackage{caption}

\title{Discrepancies in Dynamic Yield Stress Measurements of Cement Pastes
}

\author{
  Subhransu Dhar \\
  Research Unit of Building Materials and Technology,\\
  TU Wien, Karlsplatz 13, Vienna, 1040, Austria\\
  subhransu.dhar@tuwien.ac.at \\
  \And
  Teresa Liberto \\
  Research Unit of Building Materials and Technology,\\
  TU Wien, Karlsplatz 13, Vienna, 1040, Austria\\
  teresa.liberto@tuwien.ac.at \\
  \And
  Catherine Barentin  \\
  Institut Lumi\`ere Mati\`ere, Univ. de Lyon,\\
  Universit\'e Claude Bernard Lyon 1\\
  CNRS, Villeurbanne, Lyon F-69622 France\\
  catherine.barentin@univ-lyon1.fr \\
  \And
 Thibaut Divoux\\
 ENSL, CNRS, Laboratoire de physique,\\ 
 Lyon, F-69342, France\\
 thibaut.divoux@ens-lyon.fr \\
  \And
  Agathe Robisson* \\
  Research Unit of Building Materials and Technology,\\
  TU Wien, Karlsplatz 13, Vienna, 1040, Austria\\
 agathe.robisson@tuwien.ac.at \\
}

\begin{document}
\maketitle

\begin{abstract}
The dynamic yield stress associated with the flow cessation of cement pastes is measured using a rheometer equipped with various shear geometries such as vane, helical, sandblasted co-axial cylinders, and serrated parallel plates, as well as with the mini-cone spread test. Discrepancies in yield stress values are observed for cement pastes at various volume fractions, with one to two orders of magnitude difference between vane, helical and mini-cone spread measurements on the one hand, and co-axial cylinder and parallel plate measurements on the other hand. To understand this discrepancy, the flow profile of a cement paste in the parallel-plate geometry is investigated with a high-speed camera, revealing the rapid formation of an un-sheared band near the static bottom plate. The width of this band depends upon the rotational velocity of the top plate, and upon the shear time. Recalculation of shear stress shows that the reduced sheared gap alone cannot explain the low measured yield stress. Further exploration suggests the formation of zones with lower particle content, possibly linked to cement particle sedimentation. Here, we argue that the complex nature of cement pastes, composed of negatively buoyant non-Brownian particles with attractive interactions due to highly charged nano-size hydration products, accounts for their complex rheological behavior. 
\end{abstract}

% keywords can be removed

\section{Introduction}\label{sec1}

Concrete is a widely used building material composed of a binder (cementitious materials, water, additives) and fillers (aggregates, sand, etc.) \citep{thomas2009materials}. The rheological properties of concrete are largely determined by the rheology of the cementitious binder. In practice, these properties are typically specified before construction starts. This involves ensuring that the concrete can be easily mixed into a uniform slurry, the mixture remains "stable" without cement particles or aggregates segregating under gravity, and the concrete can flow when subjected to specific conditions. For instance, conventional concrete should flow when vibrated, self-compacting concrete should flow under its own weight, and concrete designed for additive manufacturing should gel when extruded from a 3D printer  \citep{roussel2012understanding}.

The rheological characterization of cement pastes has, therefore, been a focus of both practitioners and academics. However, this task is notably challenging due to the complex rheological behavior of cement, characterized as a thixotropic yield stress fluid \citep{yim2013cement,roussel2010steady,roussel2012understanding}. Its characterization is also complex, and attributed to issues such as wall slip during testing
\citep{saak2001influence} and the formation of heterogeneities, i.e., a gradient in particle concentration caused by sedimentation and migration of cement particles during flow \citep{bhatty1982sedimentation,perrot2012yield,baumert2020influence}. 

Cement paste exhibits dual characteristics: first, it is a suspension of negatively buoyant non-Brownian particles with a diameter of about $10~\rm \mu m$, and a density of approximately $3150~\rm kg/m^3$, dispersed in water at a solid volume fraction typically between 40 and 50\,\%. Second, it transforms into a colloidal gel due to the presence of highly charged nano-particles (hydration products) that precipitate at the surface of cement particles and in the interstitial fluid. 
These nano-particles are responsible for the strong electrostatic interactions occurring both between cement particle surfaces and among themselves \citep{plassard2005nanoscale,allen1987development,jonsson2005controlling,bullard2011mechanisms,goyal2021physics}.
Indeed, when cement particles are mixed with water, a dissolution/precipitation process begins. During this process, nano-particles of calcium-silicate-hydrates (C-S-H) with a size of approximately $5\,\rm nm$ are generated \citep{dal2022atomistic,allen1987development,jonsson2005controlling}. Despite their low concentration in the early stages, i.e., less than a few percent \citep{scherer2012hydration}, the strong interactions between C-S-H particles and between C-S-H and charged cement particles, facilitated by an interstitial solution of high ionic strength \citep{bullard2011mechanisms,goyal2021physics}, contribute to the rheological properties of the mixture \citep{marchesini2019irreversible}. This cohesive network builds up with time at rest \citep{hannant1985equipment,wallevik2009rheological,liberto2022small}, leading to the macroscopic yield stress of cement (and concrete) \citep{jonsson2004onset,jennings2000model,roussel2012origins}, and ultimately, its hardening \citep{goyal2021physics}. 
The yield stress increases with time at rest, even within the first hour \citep{figura2010hydrating}. When subjected to shear, this multi-scale network of micron-sized cement and nano-sized C-S-H particles is broken down (fluidized), resulting in the flow of the slurry  \citep{banfill1991rheology}. This process exhibits macroscopic reversibility over a time frame of about 1~hour, giving rise to the thixotropic and shear-thinning properties of cement \citep{lootens2004gelation,struble1995viscosity,wallevik2005thixotropic,marchesini2019irreversible,link2023mechanisms}. 

Previous works have highlighted that, under flow, heterogeneities in microstructure grow due to the migration and sedimentation of cement particles \citep{feys2018measuring, ley2019challenges, cardoso2015parallel}. The resulting particle segregation that may occur during rheological characterization can lead to misleading data. In particular, the lack of data reproducibility becomes more apparent when utilizing parallel plates \citep{cardoso2015parallel}. Despite efforts to minimize slippage by adjusting the plate roughness \citep{nehdi2004estimating}, no definitive scenario underpinning such inaccuracies has been proposed \citep{haist2020interlaboratory}.

The present work aims to shed light on the origin of these phenomena by investigating the potential development of heterogeneities in flow profile or in particle content upon shearing a cement paste. It focuses on understanding the discrepancies in yield stress measurement, depending on the rheometer geometry employed.
The investigation involved measuring the yield stress ($\tau_0$) of cement pastes prepared at various water-to-cement ratios (w/c) (or, equivalently, volume fractions $\phi$) using different rheometer geometries, namely serrated parallel plates, sand-blasted co-axial cylinders, vane, helix, and using a mini-cone spread test \citep{roussel2005fifty}.
To explore the important disparities in recorded yield stress values between vane, helix, and mini-cone spread on the one hand and co-axial cylinders and parallel plates on the other hand, a high-speed camera was set up. Recorded images of the paste under shear at the edge of the parallel-plate geometry revealed the presence of shear banding, i.e., the coexistence of an arrested band near the fixed plate and a moving band next to the rotating plate. 
We further show that shear history is actually critically contributing to the observed flow heterogeneities. Additionally, cement pastes with lower particle content (higher water-to-cement ratios) were tested, and their yield stress was compared with values observed on denser paste measured with the plate-plate geometry, providing further support for the conclusions drawn. 

The outline of the paper is as follows: in section~\ref{sec:MatMet}, we detail how the cement pastes are prepared and explicitly define how the dynamic yield stress is evaluated in all the different geometries. Section~\ref{sec:Results} reports our main results, and shows how the presence of a shear band in the parallel plate geometry impacts the determination of the yield stress. We further demonstrate how shear history affects the transient response of cement paste during steady shear experiments. Finally, we discuss the previous results in light of the literature in section~\ref{sec:Discussion}, and conclude in section~\ref{sec:Conclusion}.

\section{Materials and methods}
\label{sec:MatMet}

\subsection{Material and mixing process}

Samples were prepared with a commercial cement (Contragress, 42.5 R, $C_{3}A$-free) composed of particles (ground clinker and  3-5\% gypsum) with percentile values D10, D50 and D90 of 4, 18 and 43~$\mu$m, respectively, determined by laser diffraction using a Malvern Mastersizer 3000. The density of the cement particles, denoted as $\rho_c$, is approximately $3150~\rm kg/m^3$ considering the density of its primary constituent, tricalcium silicate \citep{thomas2009materials}. This results in a particle/suspending fluid density difference of $\Delta\rho\approx 2150~\rm kg/m^3$, assuming an interstitial solution density (ionic water) of $1000~\rm kg/m^3$. In some experiments, a polycarboxylether (MasterGlenium ACE 430 from Master Builders) superplasticizer, denoted as SP, was used and pre-mixed in water.

For yield stress measurements, samples were prepared with various water-to-cement ratios (w/c) as reported in Table~\ref{tab:Samples}. The solid volume fraction of cement particles, denoted as $\phi$, was calculated based on the recipe. For the {imaging and step-shear experiments} conducted in the parallel-plate (PP) geometry, only the sample with $\phi = 0.45$ (w/c = 0.4) without any SP was used.

\begin{table}[h!]
\setlength{\tabcolsep}{40pt}
\centering
\begin{tabular}{c c c}

\hline
 w/c &  $\phi$ &     SP \% (bwoc) \\ 
\hline
 0.35 &  0.48   & 0   \\
 0.40 &  0.45 &  0   \\
 0.40 &  0.45 &  0.3  \\
 0.45 &  0.42 &  0  \\
 0.48 &  0.40 &  0  \\
 0.52 &  0.38 &  0  \\
 0.55 &  0.37 &  0  \\
 0.57 &  0.36 &  0  \\
 \hline
\hline
\end{tabular}
\captionsetup{width=31pc}
 \caption{Recipes of cement slurries studied. w/c is water to water-to-cement ratio in mass, $\phi$ is the solid volume fraction of cement particles. SP is superplasticizer, and  bwoc stands for "by weight of cement."}
\label{tab:Samples}
\end{table}

Two different mixers were used to prepare the slurries. For large quantities (i.e., 50~mL for flow curve rheological tests and 80~mL for cone spread), an IKA\textsuperscript{\textregistered} STARVISC 200-2.5 control overhead 3 blade stirrer (model: Starvisc 200-2.5 C S000) was used. For small quantities (e.g., 10~mL for imaging experiments in PP geometry), an IKA\textsuperscript{\textregistered} ULTRA-TURRAX\textsuperscript{\textregistered} Tube Drive P control mixer was used. 
  
The mixing protocol involved a gradual addition of cement powder to water (sometimes pre-mixed with the superplasticizer), taking approximately 30\,s. Subsequently, the walls of the beaker were scraped, and the mixture was further stirred for 3 minutes at 800\,rotations per minute (rpm) in the overhead mixer and at 6000\,rpm in the ULTRA-TURRAX\textsuperscript{\textregistered} one.  Following this, the cement slurry beaker was sealed to prevent drying and allowed to rest for 10\,minutes. After the resting period, just before testing, the paste was scraped, quickly stirred with a spatula, and mixed again for 1\,minute (at the same rpm as above). 
This procedure was implemented based on a preliminary study indicating that the yield stress evolution was slower after a $10~\rm min$ rest and re-mix. Finally, at the end of the remixing process, the paste was immediately loaded into the rheometer or mini-cone for testing. The time delay between the end of mixing and the beginning of testing ranged from 60 to $90~\rm s$. 

To ensure reproducibility, only one batch of cement, stored in a sealed plastic container, was used. However,  during the testing period (ca. 18 months), aging may have caused
slight variations in measured shear stress. 

\label{subsection:Material_for_imaging}

\subsection{Rheological characterization \& imaging}

For all rheological characterization, a torque-controlled rheometer (MCR 302, Anton Paar) was used. The instrument was equipped with various geometries, including vane, helical (double helix structure), sandblasted co-axial cylinders (CC), and serrated parallel-plates (PP), as illustrated in Figure \ref{fig:Geometries_used}. The dimensions of these geometries are provided in the supplementary information. 
In the case of the PP geometry, special care was taken to spread an appropriate amount of paste on the plate, covering about 80\% of the plate surface initially, before bringing the top plate to the measuring position corresponding to a gap $H=1$\,mm. This precaution aimed to avoid excessive squeezing, which could potentially lead to particle migration, resulting in a heterogeneous paste microstructure even before the start of the shear experiment  \citep{ramachandran2010particle}. To minimize paste drying, a solvent trap, and water-filled groove were used, except during imaging experiments (see section~\ref{sec:Imaging}). Additionally, the yield stress of the pastes was measured using the mini-cone spread test \citep{roussel2005mini,roussel2006correlation,tan2017reproducible}. The dimensions of the mini-cone used can be found in the supplementary information. A constant temperature of $20~^\circ$C was maintained for all rheological experiments, while the mini-cone spread test was conducted at room temperature, i.e., approximately $22~^\circ$C.

\begin{figure}[!t]
\centering
 \includegraphics[width=16pc]{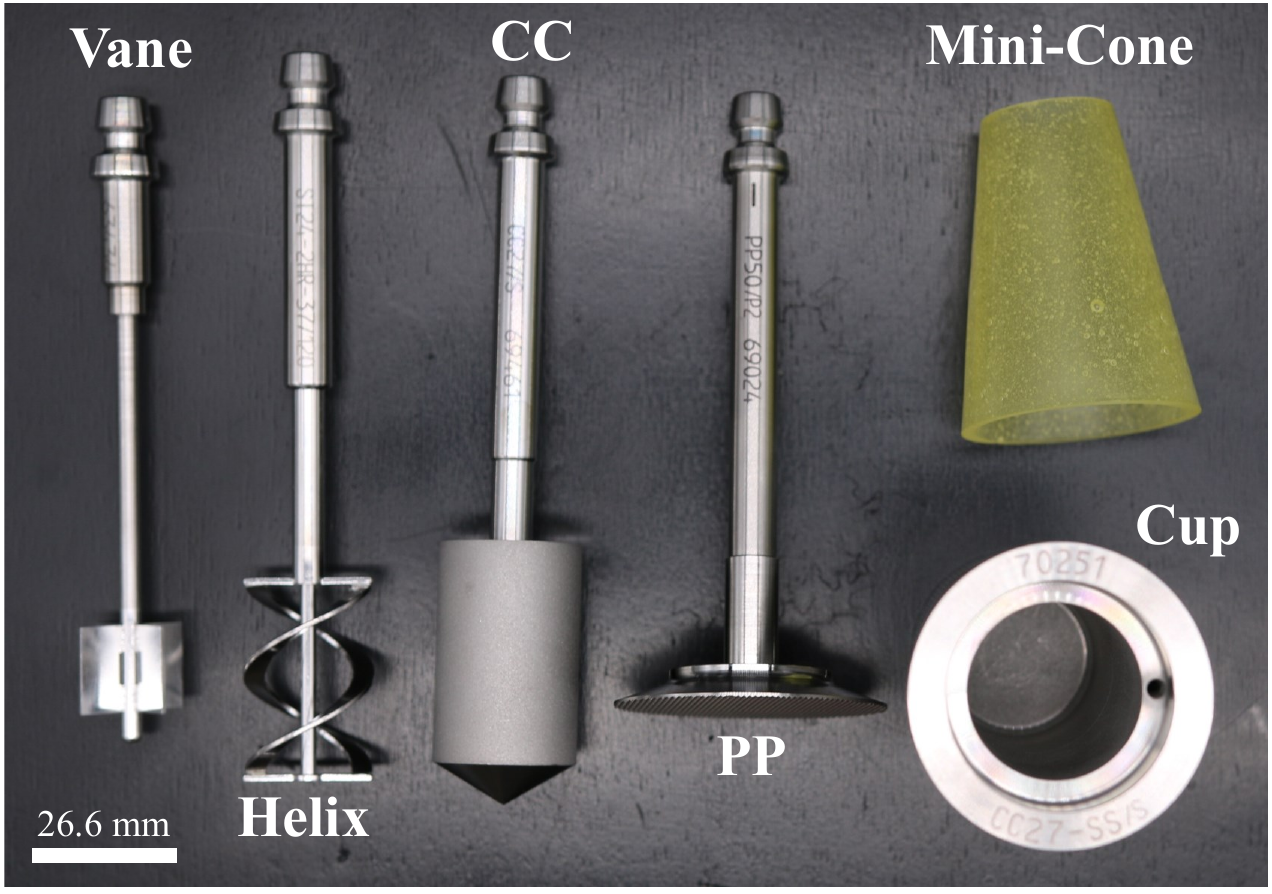}
 \captionsetup{width=31pc}
\caption{Geometries used for rheological measurements. From left to right: vane, helix, sand-blasted coaxial cylinder (CC), serrated parallel plate (PP), mini-cone, and sandblasted outer cup.}
\label{fig:Geometries_used}
\end{figure}

\subsubsection{Dynamic yield stress measurement with the rheometer}
\label{Dynamic_yield_stress}

Flow curves are recorded with the rheometer by decreasing the shear stress $\tau$ through logarithmically spaced steps, using N=25 steps per decade, with duration logarithmically increasing from 3 s to 5 s each. 
The stress range explored was adjusted for each test to cover stress values spanning one order of magnitude above and below the yield stress $\tau _{o}$ (for exact values, see Table~\ref{tab:Imposed_stress} in the appendix). 

\begin{figure}[h!]
\centering
\includegraphics[width=30pc]{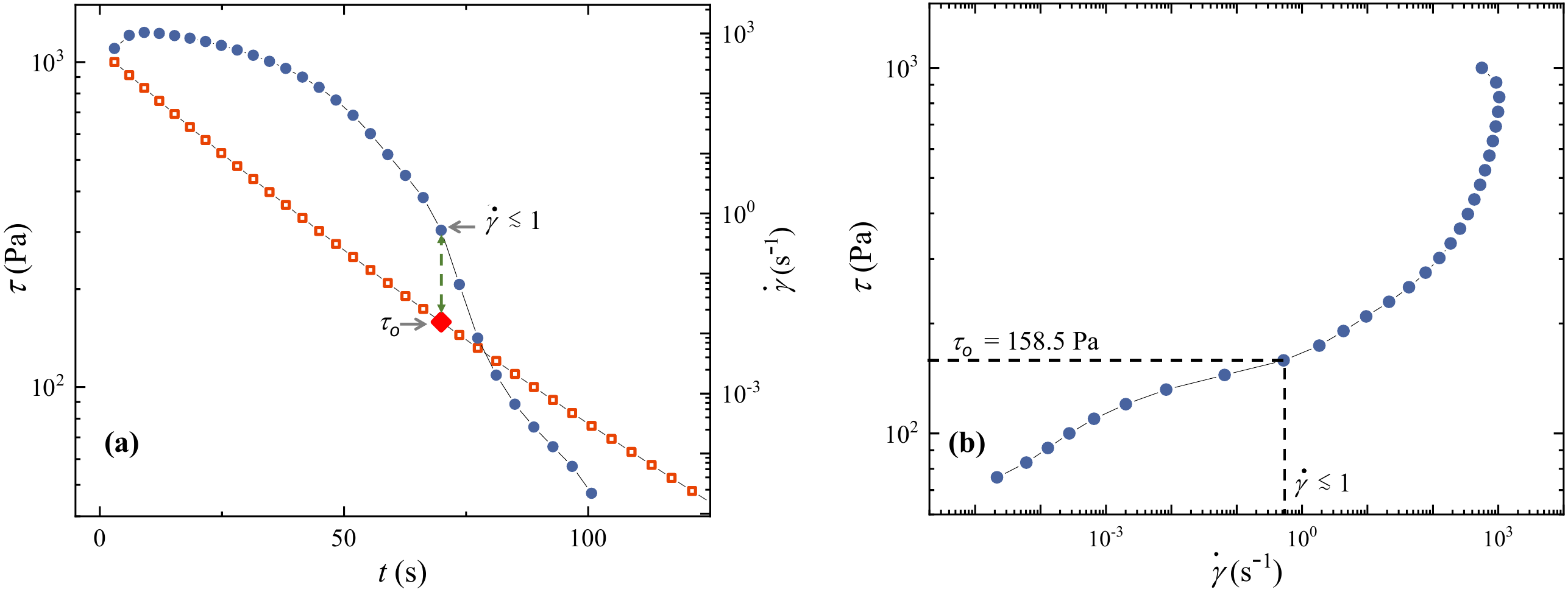}
\captionsetup{width=31pc}
\caption{(a) Imposed stress $\tau$ (orange squares) and corresponding measured shear rate $\dot \gamma$ (blue dots) as a function of time for a cement paste with $\phi$=0.45 sheared in the vane geometry. The imposed stress range was set from 1000 to 10 Pa with a logarithmically decreasing value with 25 data points per decade. The shear duration for the first step was $3~\rm s$, and logarithmically increased to 5\,s for the last step. The red diamond marks the dynamic yield stress $\tau_{o}= 158.5~\rm Pa$ defined as the value of $\tau$ at $\dot\gamma$ just below$~1~\rm s^{-1}$. (b) Imposed shear stress $\tau$ vs. shear rate $\dot \gamma$ during stress ramp down for the same test as in (a).}
\label{fig:stress_strain_time}
\end{figure}

We observe that the shear rate decreases with time for decreasing shear stresses and drops sharply for shear rates $\dot \gamma$ close to $1~\rm s^{-1}$, as illustrated in Figure \ref{fig:stress_strain_time} for a cement paste with $\phi=0.45$ sheared within the vane geometry. This point where $\dot \gamma$ $\approx$ $1~\rm s^{-1}$ also corresponds to an apparent inflection point. 
This observation prompted us to define the dynamic yield stress of the paste as $\tau_o=\tau$ at which $\dot \gamma$ crosses below 1~\rm $s^{-1}$. Moreover, this somewhat arbitrary definition of $\tau _{o}$ was supported by testing a more conventional approach. Indeed, fitting the flow curve $\tau(\dot{\gamma})$ with the Herschel-Bulkley model led to negative values of $\tau_o$ in the case of parallel plates and coaxial cylinders geometries. In these geometries, the drop in shear stress values with $\dot \gamma$ that occurs at low rotational velocities is likely linked to flow heterogeneities that develop locally and are investigated in more detail in the rest of the article. Each test was repeated twice using a freshly prepared paste.

\subsubsection{Dynamic yield stress measurement with the mini-cone spread test}
\label{sec:spreadtest}

The dynamic yield stress of the cement paste was also measured using a mini-cone spread test often used by cement and concrete end-users \citep{roussel2012understanding}. The protocol used here first entailed wetting a mini-cone (Fig.~\ref{fig:Geometries_used}) and a smooth horizontal steel surface. The cone was held tight on the table with the larger diameter resting on it, and it was then filled with cement paste to the brim. The cone was then lifted (in about 1\,s), and the paste was allowed to spread onto the surface under its own weight. The measured radius $R_S$ of the cement spread was finally used to calculate a dynamic yield stress $\tau_{o}$ using the following expression: $\tau _{o} = {225\rho g V^2}/({128 \pi^2 {R_S}^5})$ \citep{roussel2005fifty,tan2017reproducible}, where $\rho$ is the density of the cement paste, $g$ is the acceleration due to gravity, and $V$ is the paste volume contained in the mini-cone. Each test was repeated twice with a freshly prepared sample.

\subsubsection{Torque evolution with time at constant angular velocity}
\label{torque_evolution}
Constant angular velocity experiments were performed on the $\phi$ = 0.45 paste in the parallel plate, the vane, and the helix geometries. The aim of these experiments is to compare the torque evolution at various constant-imposed rotational velocities in these three geometries. In the parallel-plate geometry with a gap $H=1~\rm mm$, the velocities applied for 300~s to the upper plate were $\Omega$ = 0.06, 3.0 and $30.0~\rm rad/s$ (equivalent to the following imposed shear rates calculated for a Newtonian fluid: $\dot{\gamma}_{i}$ = 1.5, 75 and 750~s$^{-1}$ at maximum radius $R_{max}$). 
The imposed velocities applied using the vane geometry were $\Omega = 1.05, 10.47, 20.93~\rm rad/s$. Finally, with the helix geometry, the angular velocities were $\Omega = 1.05, 5.23,~ {\rm and}~10.47~\rm rad/s$. A new sample was loaded for each $\Omega$. The torque evolutions with time were compared and discussed for the three geometries.

\subsubsection{Influence of shear history in the parallel-plate geometry}
\label{shear_history}
The evolution of torque $M$ was measured on the $\phi$ = 0.45 cement paste at an imposed rotation speed $\Omega=3~\rm rad/s$ for $300~\rm s$, with no prior shear, and with $300~\rm s$ of prior shear at either 0.6 or $6~\rm rad/s$. A rest time of $60~\rm s$ was imposed between the two shear rate steps. For each test, a newly mixed cement paste was loaded in the geometry, and each test was repeated twice. The evolution of torque is then compared in the three cases. 
The evolution of shear history was also explored in the vane and helix geometries (see appendix \ref{step_shear_history}).

\subsubsection{Imaging of the paste under shear in the parallel-plate geometry}
\label{sec:Imaging}

\begin{figure}[h!]
\centering
\includegraphics[width=30pc]{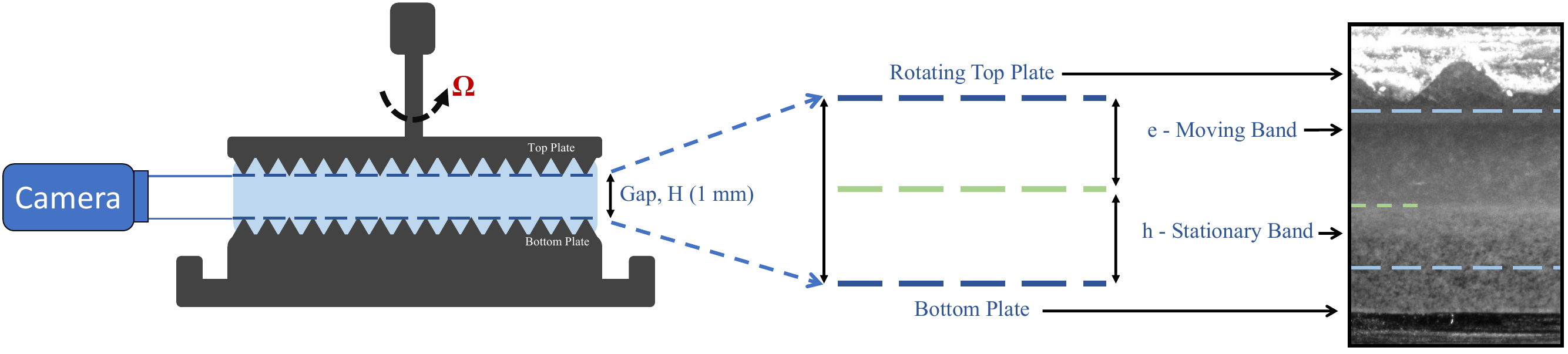}
\captionsetup{width=31pc}
\caption{Schematic of the imaging setup and width of the bands. The gap between the tip of the spikes of the plates is $H= 1~\rm mm$. The moving band corresponds to the upper band of width $e$, while the stationary band of width $h$ is in contact with the bottom plate. $\Omega$ denotes the angular velocity imposed on the top plate. On the right end, a picture of band formation is shown when an angular velocity $\Omega=100~\rm s^{-1}$ is imposed. The image is obtained by superimposing 10 consecutive images at $\approx$ 75~s from the start of shear. The tips of the spikes of the top and bottom plate are represented by deep blue dashed lines and the separation between the moving and stationary band is indicated by the green dashed line. A video of the gap during the stress ramp can be seen in supplementary information.}
\label{fig:Schematic_of_imaging}
\end{figure}	

Due to the opacity of the sample, the paste motion could only be observed on the outer edge of the sample, i.e., at the paste-air interface where $r=R_{\textrm{max}}$. The sample was illuminated with an LED fiber optic illuminator. For each rotation speed $\Omega$, images (exposure time of $50~\rm\mu$s and dimension of 900 $\times$ 700 pixels) were collected at $2~\rm Hz$ with a camera (Basler acA1920-155um). Calibration was done imaging a flexible plastic scale fixed to the edge of the plates. Image analysis was performed using ImageJ \citep{schneider2012nih}. Two distinct bands were identified: a moving band of width $e$ and a stationary band of width $h=H-e$. The term ``moving band'' pertains to the region within the gap that is actively in motion. Within this zone, the paste can either be sheared or move in a plug-flow manner. Determining the type of flow would require velocity analysis, which is beyond the scope of this work. The width of the stationary band $h$ was measured by carefully observing five consecutive images (covering a time interval of 2.5~s, small enough compared to the interval of 30/60~s between measurement points), then by measuring the gap width where pixels did not exhibit contrast variation (i.e., pixel capturing a static point location) and finally by averaging values from three different positions on the image (left, middle and right). Measurements were performed at times t = 30, 60, 120, 180, 240, and $300~\rm s$ from the start of shear. The moving band of width $e$ was calculated by subtracting $h$ from the gap width $H$.

In all above flow conditions, the Reynolds number $Re = {\rho  \Omega  H^2}/{\eta}< 0.009$,  where $\rho$ is the slurry density, $\Omega$ the angular velocity of the top plate, $H$ is the gap between the plates and $\eta$ is the slurry viscosity \citep{de2022numerical}. The low value of the Reynolds number confirms that the flow is laminar.

\section{Results}
\label{sec:Results}
\subsection {Dynamic yield stress measured with various geometries and with the mini-cone spread test}
\label{yield_stress}

\begin{figure}[t] 
\centering
\includegraphics[width=17pc]{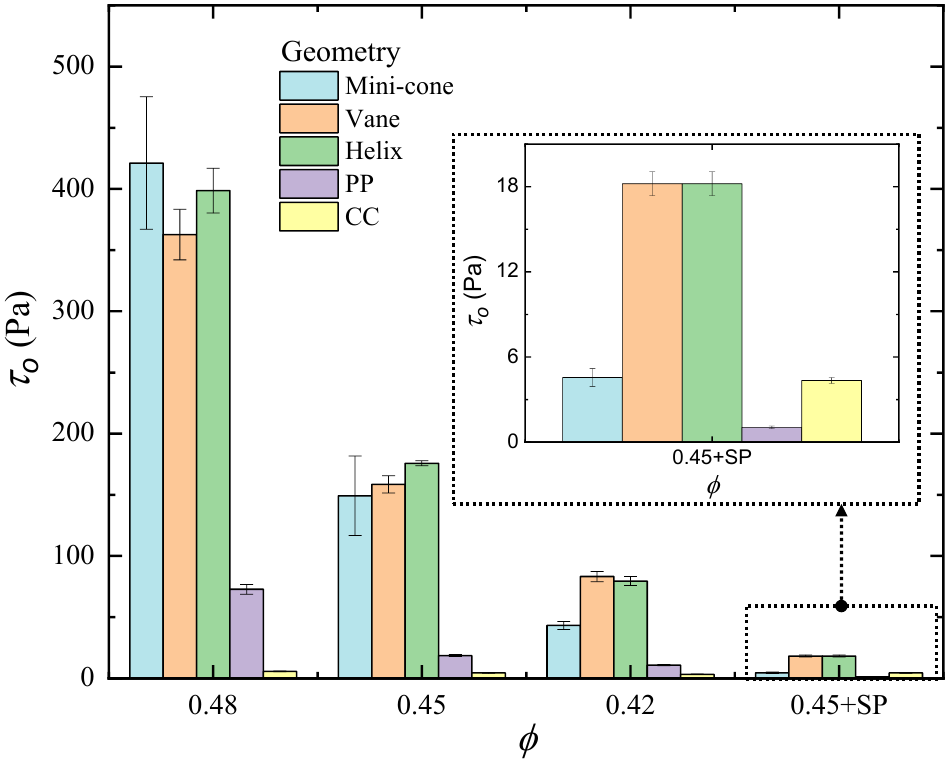}
\captionsetup{width=31pc}
\caption{Dynamic yield stress $\tau_{o}$ obtained using different geometries on cement pastes with volume fractions $\phi=0.48$,  0.45, 0.42 and 0.45 + SP (corresponding to $w/c=0.35$, 0.40, 0.45 and 0.40 + SP, respectively). Inset shows the enlarged bars for volume fraction $\phi = 0.45$ + SP. Each color represents a different geometry, with from left to right, mini-cone (blue), vane (orange), helix (green), serrated PP (purple), and sand-blasted CC (yellow). Error bars represent the range of results (maximum-minimum) measured on two independent tests.} 
\label{fig:Accumulated_YS_below_1}
\end{figure}

Figure \ref{fig:Accumulated_YS_below_1} displays the dynamic yield stress $\tau_{o}$ measured for various cement pastes with volume fractions $\phi$ of 0.42, 0.45, 0.48, and 0.45+SP. The measurements were conducted using four different geometries and using the mini-cone spread test following the method introduced in sections~\ref{Dynamic_yield_stress} and~\ref{sec:spreadtest}. The values of $\tau_{o}$ are also reported in Table~\ref{tab:Imposed_stress} of appendix \ref{appendix_stress_range}. 
A representative video of the behavior of the paste sheared within the parallel plate during the ramp-down is shown in supplementary information, and clearly illustrates the progressive formation of a shear band.

The dynamic yield stress obtained with the vane and helical geometries, and the mini-cone spread, are similar for all the volume fractions explored. In contrast, the dynamic yield stress obtained with parallel plate (PP) and co-axial cylinder (CC) geometries are significantly lower. For instance, for the cement paste with $\phi$=0.45 (corresponding to $\rm w/c = 0.40$), the average value of $\tau_{o}$ measured with the vane and helix geometries and the mini-cone spread is $\tau_o^{avg}$ = 161 $\pm$ 12 Pa. Conversely, the values of $\tau_{o}$  obtained with the parallel plates and co-axial cylinders are respectively 18.6 $\pm$ 0.9 Pa (i.e., approximately 0.12 $\tau_o^{avg}$) and  4.4 $\pm$ 0.01 Pa (i.e., approximately 0.03 $\tau_o^{avg}$). 
Note that the use of serrated parallel plates and sand-blasted cup and bob prevent wall slip, thus ruling it out as an explanation for the observed discrepancy \citep{rahman2003effect,Ballesta:2012,Isa:2007}.

\FloatBarrier 

\subsection {Width of the moving band under constant angular velocity in the PP geometry}

\begin{figure}[t!]
\centering
\includegraphics[width=0.55\linewidth]{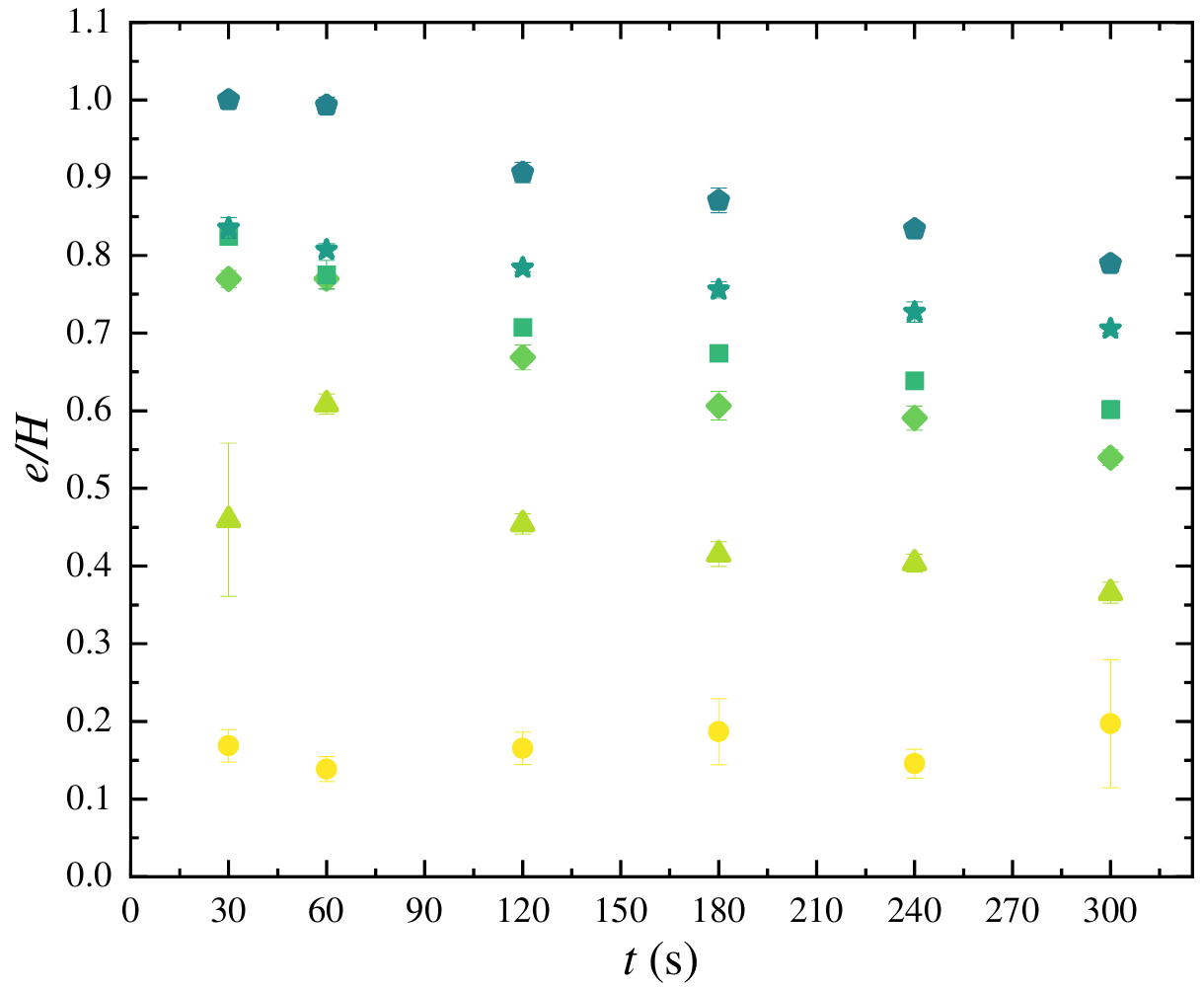}
\captionsetup{width=31pc}
\caption{Evolution of dimensionless width of moving band, denoted as $e/H$ (with $H = 1$~mm), with time for a cement paste $\phi$=0.45, for various imposed angular velocities ($\Omega$ = 0.06 -dot-, 0.60 -triangle- 3.00 -diamond-, 6.00 -square-, 18.0 -star-, $\&$ 30.0 rad/s -pentagon-) to the top plate. The colour gradient from yellow to green represents low to high angular velocities. The error bar represents the minimum and maximum value  at each data point.}
\label{fig:Shear_band_results}
\end{figure}

To understand the origin of the lower yield stress values measured with the parallel-plate geometry, we now discuss the results obtained by monitoring the motion of the paste at the outer edge of the plates. Compiled images show the existence of two distinct bands within the gap: a stationary band of width $h$ close to the bottom plate and a moving band of width $e$ near the top plate (see sketch in Fig.~\ref{fig:Schematic_of_imaging}).  The dimensionless width of moving band $e/H$, where $H=1$~mm is the gap size, is reported in Figure \ref{fig:Shear_band_results} as a function of time, for shear start-up experiments performed at fixed rotation speed $\Omega$ ranging between $0.06~\rm rad/s$ and $30~\rm rad/s$.
Results show that, within the time frame of 300~s, $e/H$ is either constant or decreases with time at all investigated rotational velocities. This result contrasts with previous observations in dense suspensions of softer particles where the shear band observed during shear start-up flow was reported to slowly grow leading to the complete fluidization of the material \citep{divoux2010transient}.

\begin{figure}
\centering
\includegraphics[width=20pc]{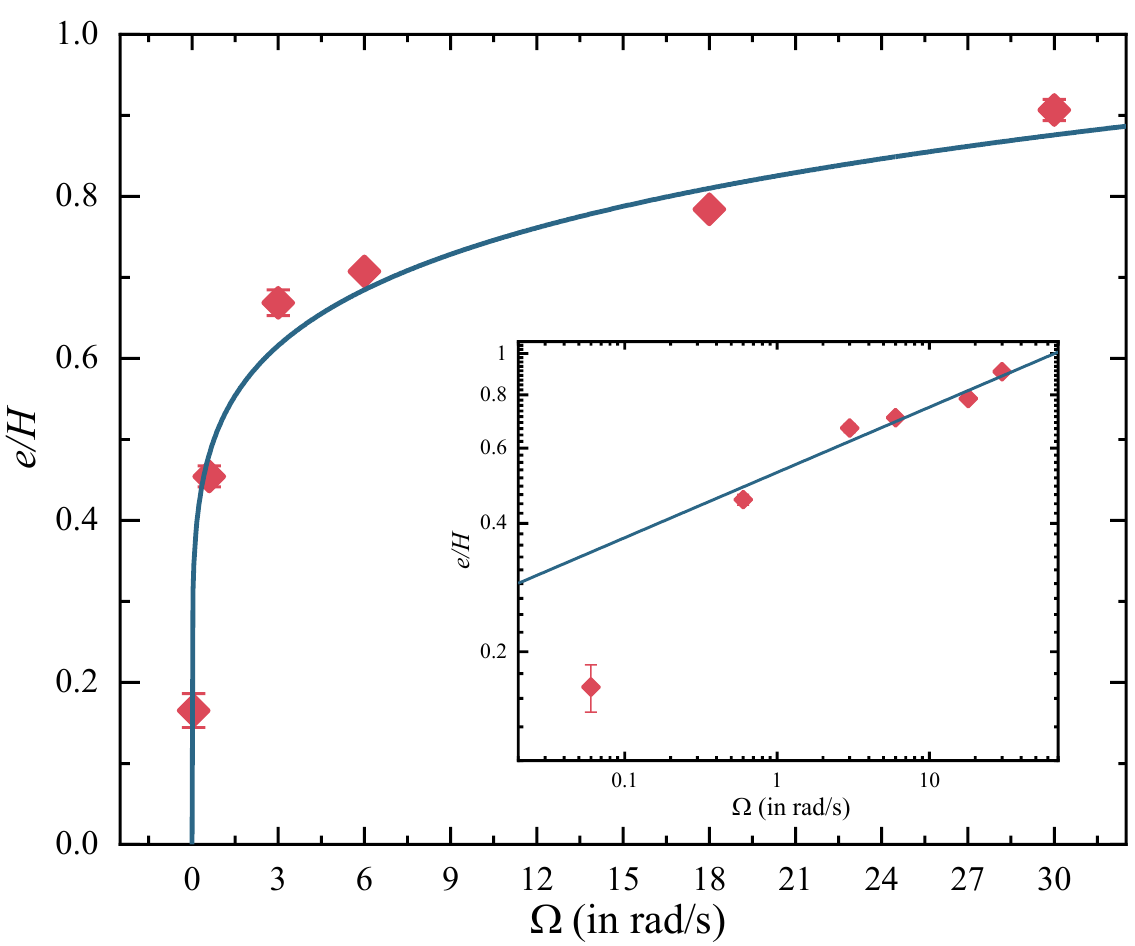}
\captionsetup{width=31pc}
\caption{Normalized width of the flowing band $e/H$ vs.~angular velocity $\Omega$, at the fixed time of 120~s after shear start up, for a cement paste $\phi$=0.45. The curve shows the best fit of the data with a power law of equation $e/H = B \times \Omega^A$, with $A = 0.15 \pm 0.03$ and $B = 0.520 \pm 0.049~\rm s^A$. Inset: same data in log-log scale. Error bars, which represent the maximum and minimum values at each data point, are of the order of the symbol size.}
\label{fig:e_vs_angular_velocity}
\end{figure}

Finally, we report in Figure~\ref{fig:e_vs_angular_velocity} the dependence of $e/H$ with $\Omega$, at a fixed time selected arbitrarily at $120~\rm s$. An increasing fraction of the gap is flowing for increasing angular velocity. The data are fitted with a power law of equation $ e/H = B\times\Omega^A$, yielding an exponent $A = 0.15\pm 0.03$. These results are in agreement with seminal observations in suspensions where sedimentation occurs, which suggests that the same mechanism is at play in the flow of cement pastes \citep{Barentin_2004, lenoble2005flow}. 

\label{subsection:Results_for_imaging}

\subsection{Evaluation of shear stress and shear rate based on the flow profile}

Building upon the local measurements presented above, the shear rate $\dot{\gamma}$ and the shear stress $\tau$ are recalculated from angular velocity $\Omega$ and torque $M$ under the assumption that the material observed within the moving band is homogeneously sheared. In other words, $e$, the width of the shear band, is used in place of the whole gap $H$.
The equations were first derived assuming that the sheared gap width is constant along the radial position $r$. Other assumptions and derivations are discussed in the appendix (\ref{radially_varying_bandwidth}).  
The shear rate $\dot{\gamma}_{e}$ at radius $R$ is given by the following expression~\citep{macosko1994rheology}:
\begin{equation} \label{eqn2_shear_rate}
	\dot{\gamma}_{e} = \frac{R\cdot\Omega}{e} =\frac{H}{e} \dot{\gamma}_{H}
	\end{equation}
The corrected shear rate $\dot{\gamma}_{e}$ changes by a factor $H/e$ from the shear rate $\dot{\gamma}_{H}$ calculated at radius $R$ using the whole gap $H$. Moreover, the shear stress $\tau_R$ at radius R can be expressed as ~\citep{macosko1994rheology}: 
%The viscosity $\eta$ of the sample can be obtained from $\tau_R$ and $\dot{\gamma}_R$ by \Cref{eqn4_viscosity}.
\begin{equation} \label{eqn3_shear_stress}
	\tau_R = \frac{M}{2\pi R^3} \Bigg[3+\diff{(\ln M)}{(\ln \dot{\gamma}_R)}\Bigg]
	\end{equation}	
where$~\dot{\gamma}_R$ is the shear rate at radius R. In practice, the factor $\diff{(\ln M)}{(\ln \dot{\gamma}_R)}$ is taken to be 1 for Newtonian fluids, i.e., $\tau_{R}=\tau_{N}={2M}/{(\pi R^3)}$ \citep{macosko1994rheology}.

\begin{figure}
\centering
\includegraphics[width=20pc]{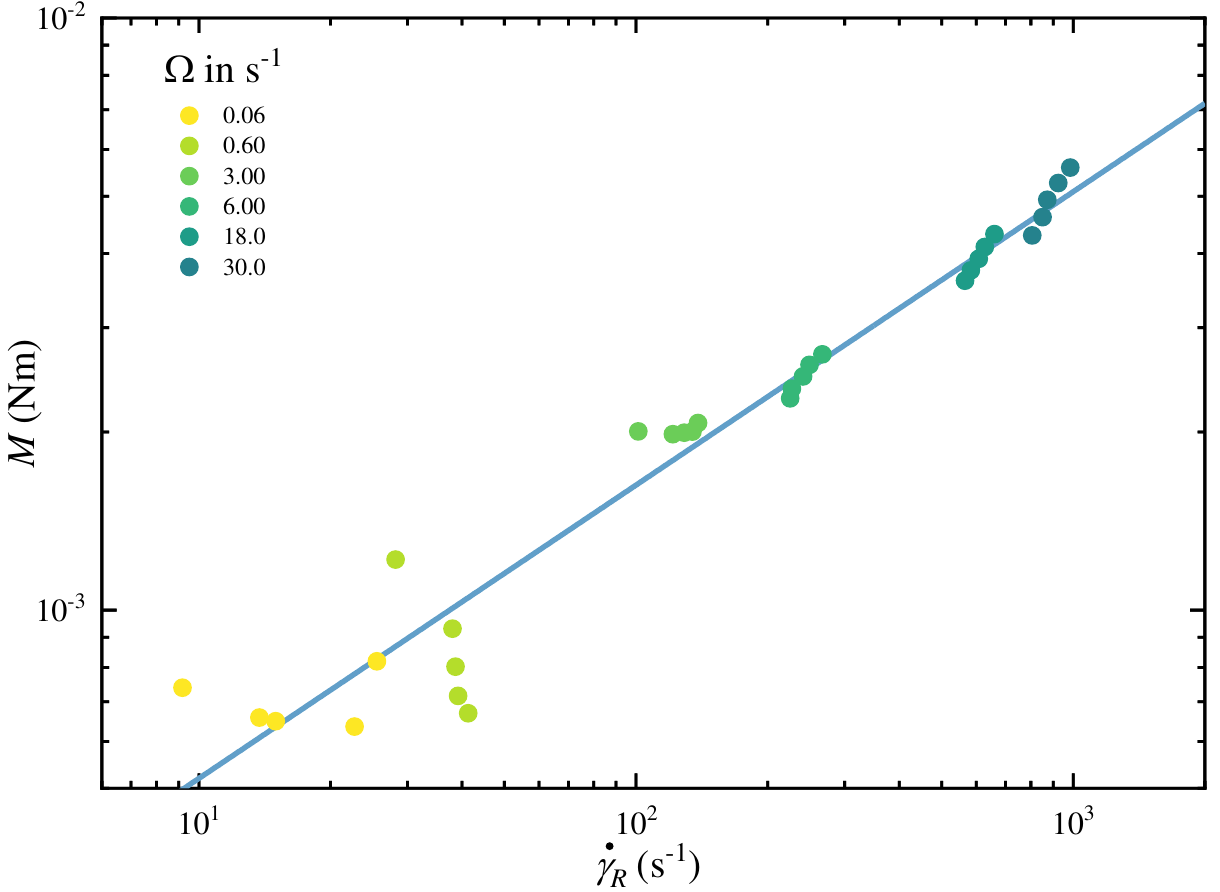}
\captionsetup{width=31pc}
\caption{Torque $M$ vs.~shear rate $\dot{\gamma}_R$ estimated at the edge of the plate for various imposed angular velocities $\Omega$. The gradient of color from light to dark implies low to high imposed angular velocity $\Omega$. Experiments were conducted on a cement paste with $\phi$=0.45 in the PP geometry. The blue line corresponds to the best linear fit of the data, $\ln~M = m\cdot  \ln~\dot{\gamma}_R + c$, with $m = 0.49\pm0.02$ and $c=-8.70\pm0.11$. }
\label{fig:slope_for_stress}
\end{figure}

In this work, the factor is \emph{a priori} unknown and can be obtained by plotting $\ln M$ vs.~$\ln \dot{\gamma}_R$ for all experiments, i.e., all imposed $\Omega$ and all measured time intervals (see Fig.~\ref{fig:slope_for_stress}). The best linear fit of the data leads to a slope value  $\diff{(\ln M)}{(\ln \dot{\gamma}_R)} =  0.49 \pm 0.02$. Using this value in eq.~\eqref{eqn3_shear_stress}, the stress $\tau$ at the edge of the plate can be calculated, leading to $\tau_R = 0.873 \tau_N$. That recalculated stress, taking into account the presence of the shear band, differs by less than a factor of $4/3$ from the stress calculated, assuming the whole gap is sheared. Therefore, the presence of a shear band cannot explain the low yield stress measured with the PP geometry, which equates to 0.115~.~$\tau _o^{avg}$, where $\tau _o^{avg}$ is the average yield stress measured with the vane, helix, and mini-cone spread test.

\subsection {Torque evolution during shear tests at imposed angular velocity }

\begin{figure*}[!t]
\centering
\includegraphics[width=31pc]{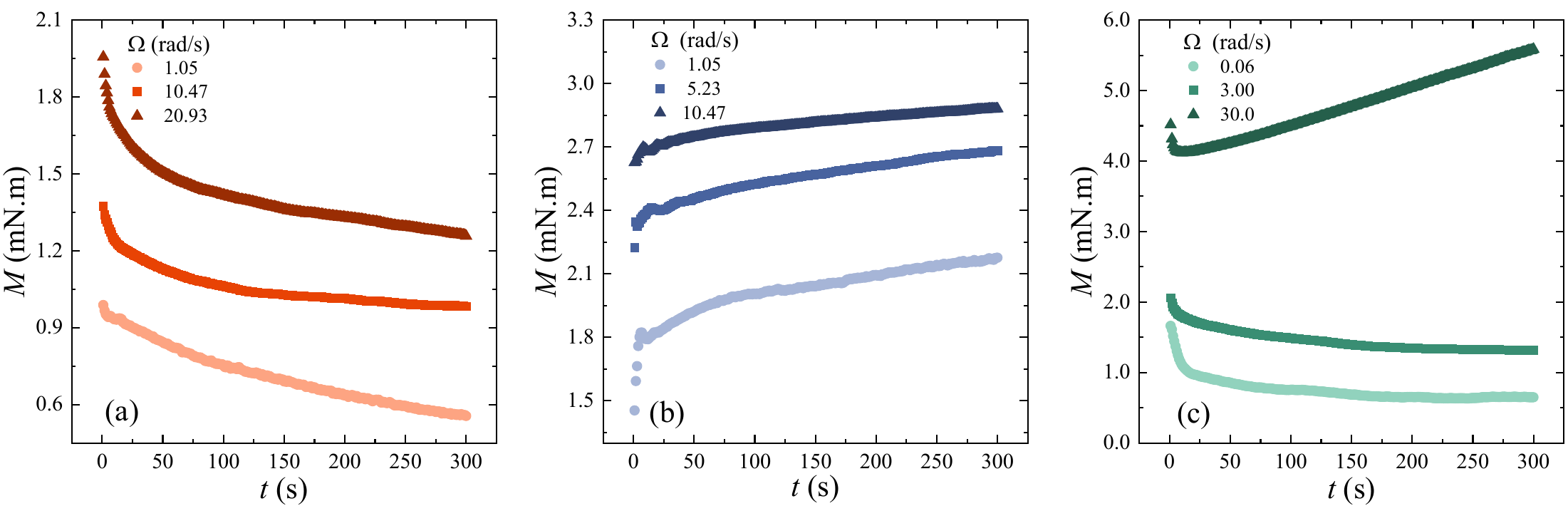}
\captionsetup{width=31pc}
\caption{Steady flow experiments at imposed rotational velocity for cement paste of $\phi$=0.45: Torque $M$ evolution as a function of time $t$ in, (a) vane, (b) helix and (c) parallel-plate geometries, for several imposed angular velocities $\Omega$. The gradient of colour from light to dark represents low to high imposed angular velocity to the geometry. }
\label{fig:Cst_Torque}
\end{figure*}

We now compare the transient torque responses observed in various geometries during steady-shear experiments. The temporal evolution of the torque on a cement paste with $\phi=0.45$ shows significant differences depending on the geometry used (Fig.~\ref{fig:Cst_Torque}). In the vane geometry (Fig.~\ref{fig:Cst_Torque}(a)), the torque decreases with time for all $\Omega$, with a maximum decrease of a factor 2 over 300~s. In contrast, the torque increases for all investigated $\Omega$ in the helix geometry (Fig.~\ref{fig:Cst_Torque}(b)), but no more than 20~\% over 300 s. Finally, in the PP geometry, the torque decreases with time for $\Omega=0.06$ and $0.6~\rm rad/s$ (to about 30-40~\% of initial torque value), and increases with time at higher angular velocities (by about 25~\% in $300~\rm s$ at $\Omega=30~\rm rad/s$). These observations, in apparent contradiction, cannot be explained by the cement slurry thixotropy alone. They suggest that there are different processes at play in the various shearing geometries. We turned to studying the impact of shear history to explore this further. 

\label{CST_Torque}

\subsection{Influence of shear history in the PP geometry}

\label{shear_step_experiments}
\begin{figure*}[!b]
\centering
\includegraphics[width=20pc]{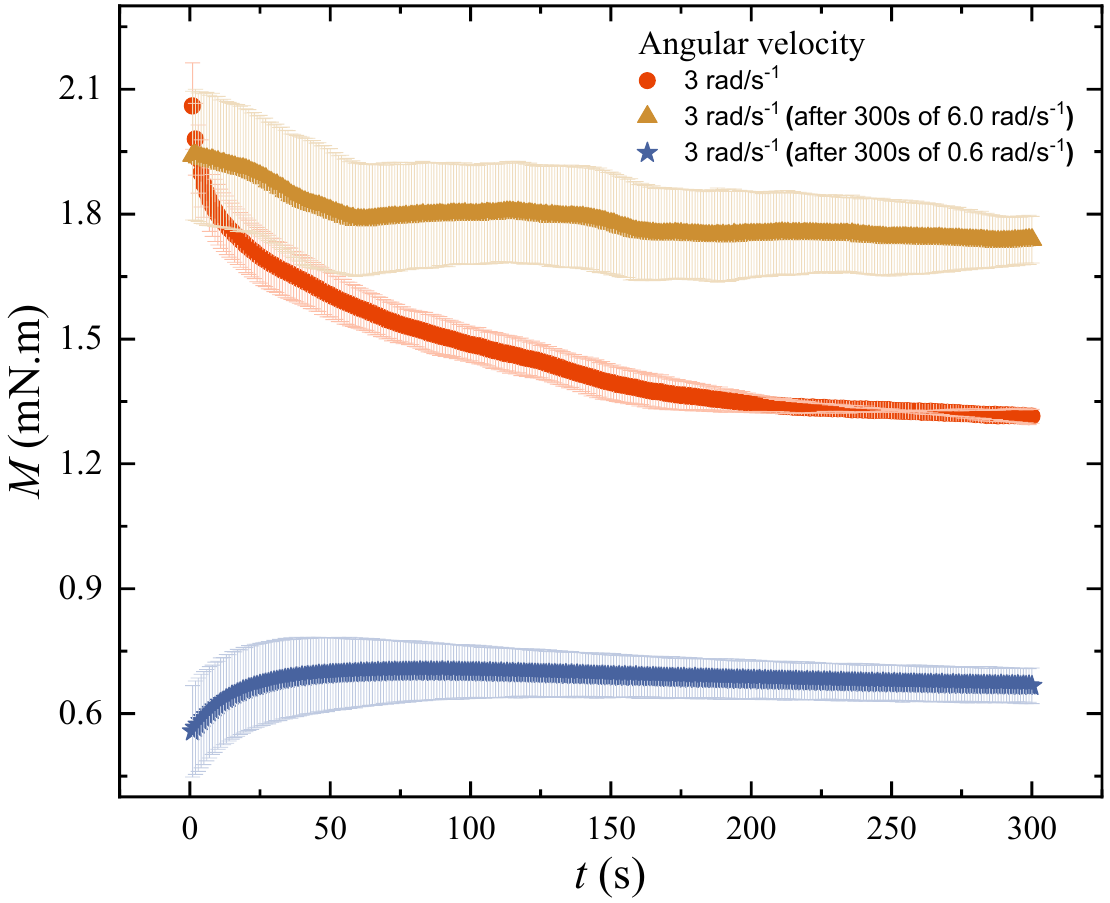}
\captionsetup{width=31pc}
\caption{Influence of shear history on the torque $M$ evolution with time $t$. Experiments were conducted at constant angular velocity $\Omega$ in the PP geometry on a cement paste with $\phi$=0.45. Red bullets: $\Omega=3~\rm rad/s$. Brown triangles: $\Omega=3~\rm rad/s$ after a first shear step at $6~\rm rad/s$. Blue stars: $\Omega=3~\rm rad/s$ after a first step at $0.6~\rm rad/s$. Error bars represented in a lighter color correspond to the minimum and maximum values taken from two tests. }
\label{fig:Step_shear_2}
\end{figure*}

Figure~\ref{fig:Step_shear_2} shows the torque response of a cement paste with $\phi = 0.45$ sheared at a constant angular velocity $\Omega=3~\rm rad/s$ in the PP geometry, after having experienced various shear histories: no previous shear, a shear at $0.6~\rm rad/s$, and a shear at $6~\rm rad/s$. 
The torque values at the end of these 3 experiments differ by a significant amount (from 0.71 to 1.79 mN.m), trends which cannot be explained by cement thixotropy alone.
In fact, when the same 2-step shear protocol is applied with the vane or the helix geometry (see Fig.~\ref{fig:Step_shear_vane_helix} in appendix), no such difference is observed: the torque difference at the end of the shear step is low (less than $\pm 15$~\%), which suggests a negligible influence of the shear history in these two geometries. Actual values at the end of each step are also shown in Table~\ref{tab:Step_shear_all} in appendix (\ref{step_shear_history}).

 \section{Discussion }
\label{sec:Discussion}

We have shown that dynamic yield stress measurements using a flow ramp-down protocol in parallel plate (PP) or coaxial cylinder (CC) geometries lead to much lower values than those obtained using the vane and helix geometries, or the cone spread. Table \ref{tab:Tau_y-Discussion} summarizes the ratio of measured $\tau_o$ with PP or CC to the average value obtained from vane, helix, and cone spread ($\tau_o^{avg}$).  The ratio spans from 1.4\,\% to 31.8\,\%, and is always higher for the PP than the CC, except for the mixture containing superplasticizer, where the PP geometry leads to lower values of $\tau_o$ than the CC geometry. 

\begingroup
\renewcommand{\arraystretch}{1.3} % Default value: 1
\begin{table}[ht]
\centering
\begin{tabular}{|p{4.5cm}||K{1.5cm}|K{1.5cm}|K{1.5cm}|K{1.5cm}|}
 \hline
 $\phi$ & 0.48 & 0.45  & 0.42 & 0.45 + SP \\
 \hline
  w/c & 0.35 & 0.40  & 0.45 & 0.40 + SP\\
  \hline  \hline
  
 $\tau_o^{avg}$ (vane, helix, mini-cone) (Pa) &394.2	&161.2	& 68.7 & 13.7	\\
  $\tau_o^{CC}$ (Pa) & 5.6	& 4.4 & 3.3 & 4.3	\\
$\tau_o^{PP} $ (Pa) &72.7	&18.6	&10.7 &1.0	\\
\hline  \hline
Ratio $\tau_o^{PP}$/$\tau_o^{avg}$ (\%) & 18.4	& 11.5	& 15.6 & 7.6	\\
Ratio $\tau_o^{CC}$/$\tau_o^{avg}$ (\%) & 1.4	& 2.7	& 4.8 & 31.8 \\
 \hline
\end{tabular}
\captionsetup{width=31pc}
 \caption{Compiled yield stress values measured with the various rheometer geometries and with the mini-cone spread test. $\phi$ is the solid volume fraction of cement particles. w/c is the water-to-cement ratio in mass. SP stands for superplasticizer. }
\label{tab:Tau_y-Discussion}
\end{table}
\endgroup
Referring to suspension rheology literature, several phenomena may account for the discrepancies highlighted above.
 First, considering that cement paste is a slurry characterized by a negative buoyancy, sedimentation of cement particles is likely during a shear test. Even though cement slurries are yield stress fluids, settling occurs when such fluids undergo shear, behaving as if the fluid had no yield stress \citep{ovarlez2012shear}.
In this case, the combined equations of Stokes-Einstein and Richardson \& Zaki \citep{miller1924stokes,richardson1954sedimentation} lead to a prediction for the sedimentation rate of approximately $1.41~\rm mm/min$ (using $\eta_{water}= 10^{-3}~\rm Pa.s$, $D50=18~\rm \mu m$, and $\phi=0.45$). Although this estimate is not precise --for instance, it overlooks non-Newtonian effects, and assumes water viscosity as an estimate of the viscosity of the interstitial fluid-- it shows that cement particles may sediment within a cement paste under shear. This particle settling can result in significant heterogeneities, particularly when the paste is sheared within the 1-mm gap of a parallel plate geometry, where the shearing planes are perpendicular to gravity. 
Simultaneously, the phenomenon of re-suspension, where the shear applied to the material is sufficient to erode a sediment layer and put back particles in suspension, can occur even at low Reynolds number   \citep{lenoble2005flow,leighton1986viscous}. In a PP geometry, while a sedimenting suspension leads to a torque decrease \citep{de2022numerical}, re-suspension leads to a torque increase \citep{lenoble2005flow}. 

Second, radial migration of cement particles is likely to occur.
Migration induced by shear has been largely discussed in buoyant suspensions and can lead to misleading results during rheological tests \citep{feys2018measuring,ley2019challenges,cardoso2015parallel,sarabian2019fully}. 
Typically, particles migrate from regions of high shear to regions of low shear, a phenomenon often linked with a torque decrease in rheological measurements \citep{merhi2005particle}. When suspensions undergo shearing within a parallel plate (PP) geometry, the situation becomes more intricate. Streamline curvature induces particle migration towards the outer edge of the plates, where the shear rate is higher, countering the migration towards the central region, where the shear rate is lower \citep{kim2008numerical}. 
Density differences between the particles and fluids can introduce additional effects due to secondary flows between the plates. In the case of negatively buoyant particles, accumulation near the bottom plate and concentration towards the plate center have been observed during rotation of the upper plate \citep{kim2008numerical,de2022numerical}, with migration depending on particle size \citep{Barentin_2004}. 
In the PP geometry, outwards migration has been associated with a torque increase \citep{merhi2005particle}, while inwards migration is associated with a torque decrease \citep{de2022numerical}. 
Moreover, the existence of heterogeneities in particle distribution even before the shear measurement starts has also been discussed. This phenomenon has been attributed to squeeze flow that may occur when the rheometer top plate comes down and applies pressure to the sample \citep{ramachandran2010particle,chow1994shear}. In such cases, a higher torque drop upon shearing start was observed. 
Although the influence of migration under flow has been modeled and discussed in the context of actual cement slurries  \citep{baumert2020influence,izadi2021squeeze}, migration was not directly experimentally observed, to the best of our knowledge. 

Third, the thixotropy of cement paste, characterized by fluidization and coagulation processes, as well as its reactivity, contributes to variations in torque over time during a shear test. The continuous precipitation of cement hydration products (e.g., nano-hydrates of C-S-H) also results in the growth of the microstructure, leading to an increase in paste viscosity \citep{ley2021resting, jennings2000model, marchesini2019irreversible, link2023mechanisms} and yield stress \citep{kabashi2023shear}.
 In the absence of the development of microstructure heterogeneities linked to migration or sedimentation, or flow heterogeneities linked to shear localization, the viscosity of a cement paste (or its torque response to a prescribed rotational velocity) reflects a dynamic equilibrium between its fluidization (the breakdown of flocs) and its structuration (coagulation driven by attractive forces between particles)\citep{banfill1991rheology,bullard2011mechanisms,goyal2021physics,jonsson2004onset,jennings2000model,roussel2012origins,liberto2022detecting}.

In the present work, no measurement of particle concentration was performed, neither through the gap thickness nor the radius. Nevertheless, the discrepancies observed in yield stress measurements and torque evolutions during steady-shear can qualitatively but consistently be explained by the processes mentioned above.
The torque evolutions recorded in the PP, vane, and helix geometries show significant differences (Figure \ref{fig:Cst_Torque}). In the helix geometry, a consistent increase in torque is observed in all cases, aligning with paste structuration. In the vane geometry, the torque decreases at all three rotational velocities, pointing towards a radial migration that would create a thin layer devoid of particles in the vicinity of the blade trajectory. This aligns with findings reported in \citep{ovarlez2011flows}, and contributes to the observed torque decrease. 
In the PP geometry, several effects are likely intertwined, making it challenging to provide a straightforward microscopic scenario. At low rotational velocities ($0.06~\rm rad/s$), a sharp torque drop with time is followed by a slower decrease. The sharp drop is likely due to the rapid formation of a thin layer near the top plate where shear is localized. This top layer is likely less rich in particles, and rapid deflocculation occurred, as reported in \citep{jarny2005rheological}.
 At longer times, additional sedimentation and/or radial migration may lead to further torque decrease. Conversely, at higher rotational velocities ($30~\rm rad/s$), a rapid torque drop is followed by a gradual increase. In this case, the highly deflocculated layer near the top plate, likely less rich in particles, may be disrupted by re-suspension at these high rotational velocities.

A side-by-side examination of torque evolutions  (Figure \ref{fig:Cst_Torque}) and band width changes (Figure \ref{fig:Shear_band_results}) over time, measured at $\Omega = 0.06~\rm rad/s$ in the PP geometry, shows that the torque keeps decreasing with time even when the width of the moving band $e$ remains constant (between 30 and 300\,s). This suggests that the material within the sheared band may be undergoing changes, either experiencing fluidization or further segregation. 
This hypothesis seems to be confirmed by the results from Figure \ref{fig:Step_shear_2}, where shear history was explored using the PP geometry. The sample previously sheared $6~\rm rad/s$ shows a higher torque response than the sample with no previous shear history, while the sample previously sheared $0.6~\rm rad/s$ shows a much lower torque. These results can, here again, be explained by the formation of a layer less rich in particles near the top plate, where strain localization occurs during shearing at $0.6~\rm rad/s$, and this layer is not disrupted at the higher shear rate of $3~\rm rad/s$.  
Conversely, a first shearing at $6~\rm rad/s$ yields the highest torque among the three tests, indicating that the shear rate is enough to counteract sedimentation by re-suspension and/or streamline curvature-induced migration. This, coupled with the continuous precipitation of hydration products, results in higher torque values during the second step when the sample is sheared at $3~\rm rad/s$. To further underscore this point, the impact of shear history on samples tested in the vane and the helix geometries is presented in the appendix (see Fig.~\ref{fig:Step_shear_vane_helix}). In these two geometries, the torque stabilizes at similar values for all shearing histories, indicating that no localized shearing appears to disrupt the measurement, likely due to 3-dimensional flows \citep{nazari2013three}.

Yield stress measurements in the CC geometry are also consistent with the above discussion. A layer richer in interstitial fluid likely forms rapidly near the bob/suspension interface, where yielding is expected to occur, localizing the strain \citep{coussot2014yield,Divoux:2016}. Here, contrary to the PP geometry, re-suspension or secondary flows do not counteract migration, resulting in a consistently low torque and, consequently, a lower apparent yield stress. 
Only the paste $\phi = 0.45$ + SP shows a higher $\tau_o$ when measured with CC compared to PP. The presence of superplasticizer leads to an important yield stress decrease, favoring sedimentation in the PP geometry, further promoting shear localization and diluting this band even further. Sedimentation is less critical in the CC geometry, where the shear planes are aligned with gravity.

\begin{figure}[!t]
\centering
\includegraphics[width=18pc]{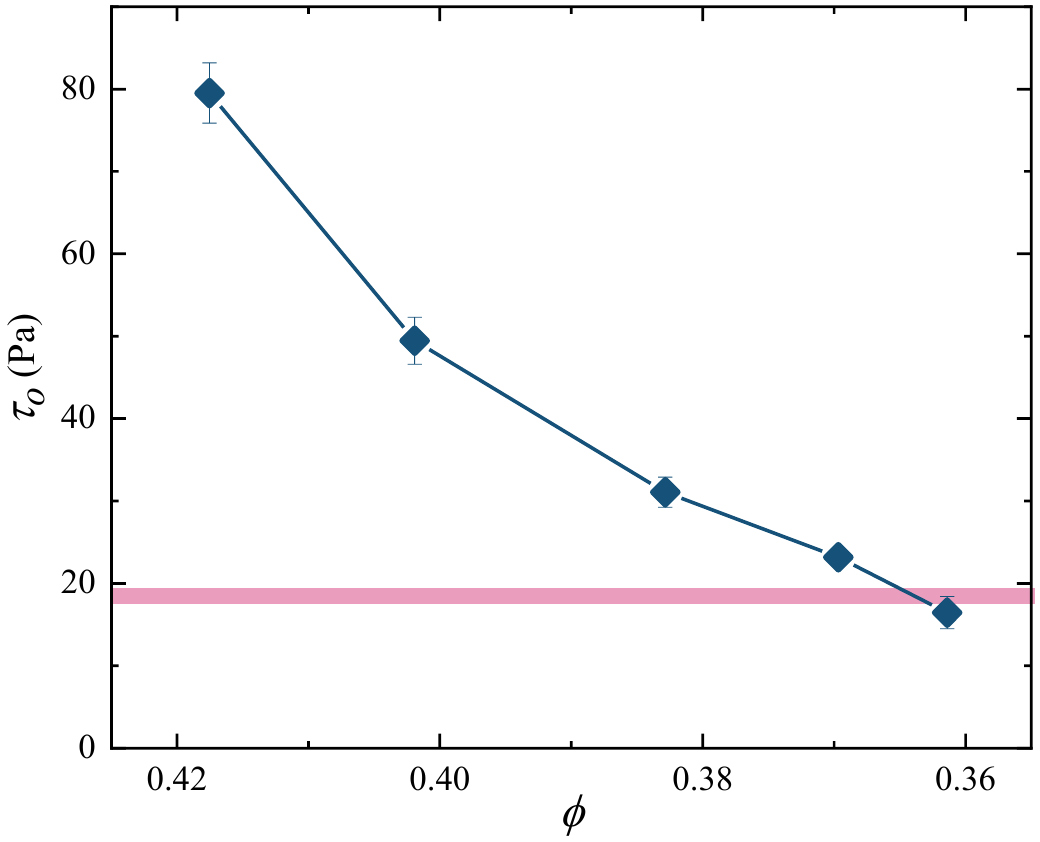}
\captionsetup{width=31pc}
\caption{Dynamic yield stress $\tau_{o}$ vs.~cement paste volume fraction $\phi$ measured using the helix geometry. The error bars are determined from two independent experiments and represent the range, i.e., maximum and minimum values, and is sometimes within the size of the symbol. The pink line highlights the value of $\tau_{o}$ measured with PP geometry and a paste of $\phi=45\%$; its width represents the range of values.}
\label{fig:yield_stress_helix}
\end{figure}

Finally, the composition of the layer where shear localizes within the parallel plates, assumed to be less dense in particle content, was determined. To do so, cement pastes with lower contents of particles, i.e., higher water contents, were prepared (see Table~\ref{tab:Samples}), and their yield stress was measured using the helix geometry, where 3D flow linked to the rotating helix is assumed to prevent the formation of a zone of localized shear.
By using the value of $\tau_{o}= 18.6 \pm 0.9~\rm Pa$ (see the pink band in Fig.~\ref{fig:yield_stress_helix}) measured on the cement paste with $\phi=0.45$ in the parallel-plate geometry, we identify that it corresponds to a yield stress of a $\phi = 0.36$ paste, measured with the helix geometry. 

\section{Conclusion}
\label{sec:Conclusion}

In this work, the dynamic yield stress $\tau _{o}$ of cement pastes prepared at volume fractions $\phi$= 0.42, 0.45, 0.48 and 0.45+superplasticizer was measured using a rheometer equipped with a vane, a helical, a sandblasted co-axial cylinder and serrated parallel plate geometry. A decreasing shear stress ramp protocol was used, and $\tau _{o}$ was defined as the stress at which $\dot \gamma$ reaches $1~\rm s^{-1}$. The yield stress was also measured using the mini-cone spread test, where the diameter of the paste spread under its own weight and the paste density were used to calculate $\tau _{o}$.  

One to two orders of magnitude differences in yield stress values were found between the vane, helical and mini-cone spread measurements on the one hand and the co-axial cylinder and parallel plate measurements on the other hand. The lower values measured with the parallel-plate geometry were further examined by imaging the cement flow within the gap using a fast-rate camera. Results highlighted a band near the static bottom plate that was not sheared, the width of which depended upon the imposed rotational velocity of the upper plate, and the time from test start. Stress and strain rate recalculation using the moving band width could not explain the lower yield stress measured in the parallel plate geometry, pointing towards the creation of heterogeneities in composition within the sheared cement paste. 

To further investigate the above results, experiments at constant imposed rotational velocity were performed on the cement paste at $\phi$=0.45. A stark contrast in torque evolution with time was observed between the vane, helix and parallel plate geometries, consistent with yield stress measurements. It can be explained by the creation of heterogeneities in flow, clearly shown by images of the sample within the gap between the parallel plates during shear. These heterogeneities in flow are likely associated with the development of heterogeneities in composition and state of flocculation that come from shear-induced migration in the concentric cylinder geometry and from shear-induced sedimentation in the parallel-plate geometry.

The study of shear history in these three geometries leads to observations consistent with the above results. At the tested velocity of 3 rad/s, 3-dimensional flows that occur in the helix and the vane geometries are enough to disrupt the potential microstructural heterogeneities created during a first shearing at low rotational velocity. On the other side, it is not enough to disrupt the coagulated layer of cement paste that formed within the parallel plate geometry.  
Finally, we show that the composition of the sheared layer within the PP geometry, starting from a paste with composition $\phi$~=~0.45, corresponds to the measured torque of a paste of lower particle content, i.e., $\phi$~$\approx$~0.36.
This study sheds light on the intricacies of rheological measurements in cement pastes, emphasizing the critical importance of selecting the appropriate geometry for characterizing their flow behavior.

\section{Appendix}

\subsection{Stress range for yield stress ($\tau_o$) measurement using various geometries}
\label{appendix_stress_range}

Table~\ref{tab:Imposed_stress} shows the imposed stress range used to measure the dynamic yield stress of the cement paste (see section \ref{Dynamic_yield_stress}).

\begin{table}[h!]
\centering
\begin{tabular}{ |c|c|c|c|c| } 

\hline
\multirow{3}{4em}{w/c ratio} & \multirow{3}{3.5em}{Geometry} & \multirow{3}{5em}{Initial stress (Pa)}   & \multirow{3}{5em}{Final stress (Pa)} & \multirow{3}{5em}{Yield stress $\tau_o$ (Pa)}  \\
 & & & &  \\
& & & &  \\
\hline
0.35 & mini-cone & - & - & 421 $\pm$ 54\\
& vane  & 2000 & 10 & 362 $\pm$ 20\\
 & helix & 2000 & 50  & 398 $\pm$ 18\\ 
 & PP & 500 & 0.1  & 73 $\pm$ 4\\ 
 & CC & 300 & 10  & 5.6 $\pm$ 0.3\\
 \hline
0.40 & mini-cone & - & - & 149 $\pm$ 32\\
& vane & 1000 & 10  &  166 $\pm$ 7\\
 & helix & 1000 & 10  & 176 $\pm$ 2\\
 & PP & 300 & 1 & 18.6 $\pm$ 0.9\\
 & CC & 130 & 10  & 4.7 $\pm$ 0.3\\
\hline
0.45 & mini-cone & - & - & 43.3 $\pm$ 3 \\
 & vane & 1000 & 10  & 87.2 $\pm$ 4.1\\
 & helix & 1000 & 10  & 79.5 $\pm$ 3.7\\
 & PP & 300 & 0.1  & 10.7 $\pm$ 0.5\\
 & CC & 60 & 1  & 3.3 $\pm$ 0.2\\
\hline
0.40 + SP & mini-cone & - & - & 4.6 $\pm$ 0.6\\ 
 & vane & 1000 & 10 & 18.2 $\pm$ 0.9 \\
 & helix & 1000 & 10 & 18.2 $\pm$ 0.9\\
 & PP & 300 & 0.1  & 1.04 $\pm$ 0.10\\
 & CC & 30 & 0.1 & 4.4 $\pm$ 0.2\\
\hline
\end{tabular}
\captionsetup{width=31pc}
 \caption{Range of shear stress $\tau$ imposed on different geometries during yield stress measurement and the measured yield stress $\tau_o$ is shown in this table. Here, $\tau = \tau_N = 2M/(\pi R^3)$ is used for parallel plate (PP) geometry; and the expression of $\tau$ for mini-cone spread, coaxial cylinder (CC), vane, and helix geometries remains consistent with the rest of the work.
 }
\label{tab:Imposed_stress}
\end{table}

\subsection{Recalculation of stress and strain rate within the parallel-plate geometry: effect of radially varying band width }
\label{radially_varying_bandwidth}

The variation of the local stress from the edge of the plates to the center is analytically computed below. The torque recorded by the rheometer is due to the sum of the local stresses in the moving gap, and can be written as eq.~\eqref{eqn_torque_balance} \citep{macosko1994rheology}:

\begin{equation} \label{eqn_torque_balance}
	M = \int_{0}^{R} 2\pi r^2 \tau(r) dr
	\end{equation}

Here, $r$ represents a radial position between 0 and maximum radius $R$, and $\tau(r)$ is the stress at a distance $r$ from the center. 
For a Newtonian fluid, $\tau(r) = \eta  \dot{\gamma}(r)$ with $\dot{\gamma}(r) = r \cdot \Omega/e$. Putting these values in eq.~\eqref{eqn_torque_balance}, the stress at the plate edge is found equal to $\tau_{N}=\frac{2M}{\pi R^3}$.
For a non-Newtonian fluid, the relation between $\tau(r)$  and $\dot{\gamma}(r)$ is different, and is evaluated here for two cases:

 \textbf{Case A:} linear relation between $\tau$ and $\eta$  where the viscosity depends upon the solid volume fraction $\phi$:  $\tau = \eta (\phi)  \dot{\gamma}$

 \textbf{Case B:} Herschel-Bulkley relation:  $\tau = \tau_{o} +  K \dot{\gamma}^n$; where $\tau_{o}$ and $K$ depend upon  $\phi$.\\

Regarding the shear band, two cases are considered here as well:

  \textbf{Case 1:} The shear band width $e$ is assumed constant with the radius $r$.
   \begin{subequations}
    \begin{eqnarray}
      \dot{\gamma} (r,z) = \frac{r \Omega}{e}~ ; ~ \text{for} ~ h-e>z>h \\
      \dot{\gamma} (r,z) = 0~ ; ~ \text{for} ~ 0< z < h-e
    \end{eqnarray}
    \label{eqn7}
  \end{subequations}

  \textbf{Case 2:}  The width of the shear band  $z(r)$ changes with the radius, i.e., it is a function of radial position $r$. Experiments have shown that $e \propto \Omega^{0.2}$, hence we make the assumption that 
 $z(r) = e (r/R)^{0.2}$.\\

 Based on the assumptions on the form of viscosity and the band width, four different cases can be considered: A1, A2, B1, and B2. 
 
\hfill \break
 \textbf{Case A1:}
    It is the simplest scenario: the band width is independent of the radius and the viscosity depends upon $\phi$ only. 

    Then eq.~\eqref{eqn_torque_balance} returns 
    
    \begin{align}
    M = \frac{\pi}{2} R^{3} \Bigg[\eta (\phi) \frac{ R \Omega}{e} \Bigg]  \nonumber 
  = \frac{\pi}{2} R^{3} \tau _{max} 
   \end{align}
   
since $R \eta (\phi) \Omega/e$ is the stress $\tau _{max}$ at $R$.  This is the same relation as that of a Newtonian fluid. Hence, under the simplest scenario, no distinctive change in stress values is present within the moving gap. 
    
\hfill \break
 \textbf{ Case A2: } Here, the band width depends on the radius $r$ and the viscosity depends on $\phi$. The equations considered are therefore $\tau = \eta (\phi) \dot{\gamma}$ and $z(r) = e\cdot (r/R)^{0.2}$.The stress becomes $\tau (r) = \eta (\phi) \Omega \cdot R^{0.2} r^{0.8}/e$. Replacing these values in eq.~\eqref{eqn_torque_balance} gives
    
     \begin{align}
 M = \frac{2\pi}{3.8} R_{max}^{3} \tau _{max} 
   \end{align}
    
    It clearly shows that there is a very small difference from that value obtained in case A1. The pre-factor in this case is $2\pi/3.8$ instead of $2\pi/4$. 
    
 \hfill \break
   \textbf{Case B1:}
    Here, the sheared band is constant within the moving gap, i.e., independent of the radius, and the fluid is now a Hershel-Bulkley fluid, i.e., $\tau = \tau_{o} +  K \dot{\gamma}^n$ and $\dot{\gamma}(r) = r  \Omega/e$. This leads to eq.~\eqref{eqn_torque_balance} taking the form:
    
    \begin{align}
 M =  \frac{2\pi R^3}{3} \Bigg[\tau_{o} + \frac{3}{n+3} K  \dot{\gamma}_{max}^n  \Bigg]
   \end{align}
    where $\gamma_{max}$ is the shear rate at the maximum radius, R.
     If we compare the calculation with that of a Newtonian fluid, giving a factor $(3/4)^n$.

   The stress is, in this case: 
   \begin{equation} \label{rheometer_equation}
	\tau_{max} = \frac{M}{\frac{2 \pi R_{max}^3}{4}} =  \frac{4}{3} \Bigg[\tau_{o} + \frac{3}{n+3} K \dot{\gamma}_{max}^n  \Bigg]\nonumber
	\end{equation}
   And K is modified by a factor $4/(n+3)$.
 
\hfill \break
  \textbf{ Case B2:}
    Finally, here is the most complex case: shear band width $z(r) = e (r/R_{max})^{0.2}$ and
    $\tau = \tau_{o} +  K \dot{\gamma}^n$, with $\dot{\gamma}(r) = r \Omega/z(r)$. This gives:
    \begin{align}
M =  \frac{2\pi R_{max}^3}{3} \Bigg[\tau_{o} + \frac{3}{0.8n+3} K  \dot{\gamma}_{max}^n  \Bigg]
\end{align}
 This result in a factor of ${4}/{3}$ in front of $\tau_{o}$. 

In all four cases, the error made by standard calculations, which assumes a Newtonian fluid with homogeneous flow within the whole gap, is at most a factor $4/3$. This cannot explain the measured difference between the measurements made with a vane and helix and that of PP geometry (up to a factor $14$). 

\subsection{Influence of shear history when shear is applied using the vane and helix geometries}
\label{step_shear_history}

\begin{figure*}
\centering
\includegraphics[width=30pc]{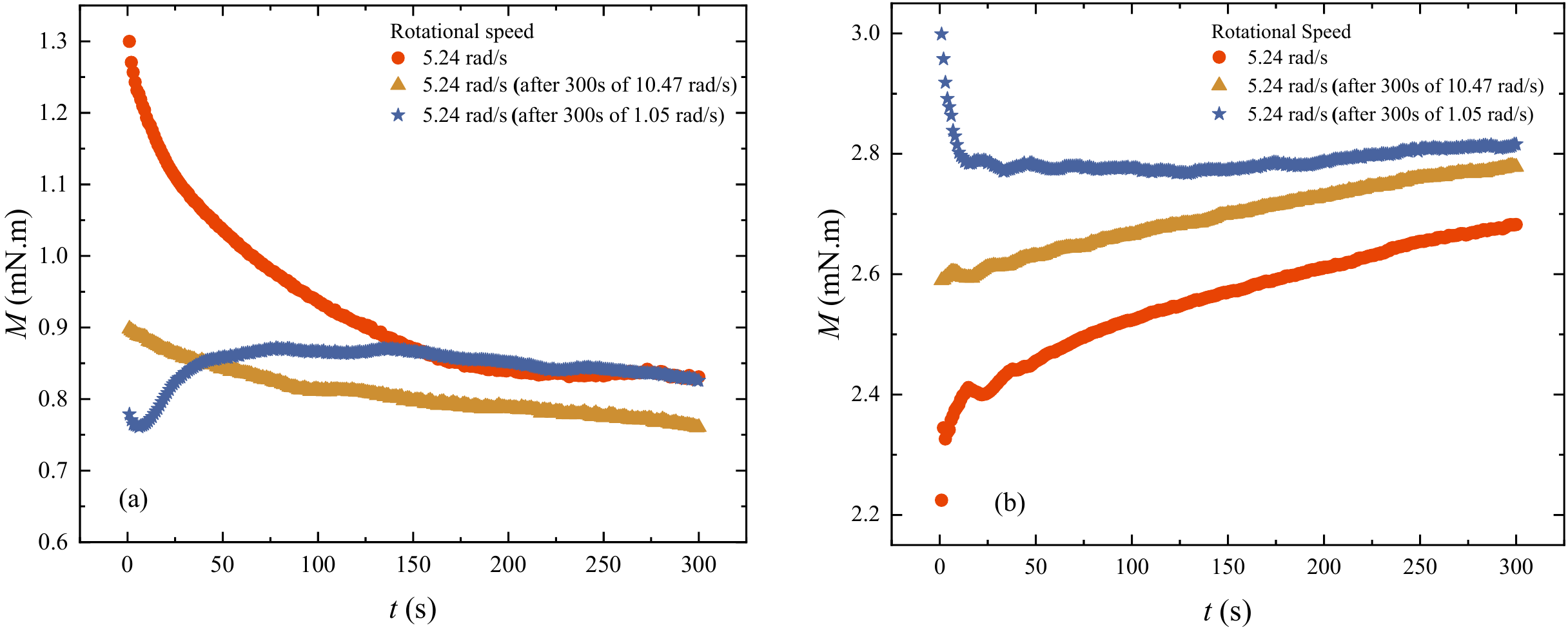}
\captionsetup{width=31pc}
\caption{Torque $M$ vs.~time $t$ at the same imposed angular velocity $\Omega$ and for 3 different shear histories, using (a) a vane and (b) a helix geometry on a cement paste with $\phi= 0.45$. Here we use a first step at $\Omega = 10.47~\rm rad/s$ for 300~s followed by $\Omega = 5.24~\rm rad/s$ (red dots) or $\Omega = 5.24~\rm rad/s$ (brown triangles). Conversely, we used a first step at $\Omega = 1.05~\rm rad/s$ for 300~s followed by $\Omega =5.24~\rm rad/s$ for 300~s (blue stars).}
\label{fig:Step_shear_vane_helix}
\end{figure*}

In this section, we discuss the shear history effects on the cement paste with $\phi = 0.45$ using the vane and helix geometries.  Three shear histories are applied. The paste is sheared at an imposed rotation speed $\Omega=5.24~\rm rad/s$ for $300~\rm s$, with no prior shear, and with $300~\rm s$ of prior shear at either $1.05~\rm rad/s$  or 10.47~rad/s. A rest time of $60~\rm s$ was imposed between the two shear rate steps. For each test, a newly mixed cement paste was loaded in the geometry. The evolution of torque is then compared in the three cases.
In this set of experiments, we compare the value of $M$ at the end of experiment, i.e., at the end of shearing at $\Omega$ = 5.24 rad/s in all three cases. Table~\ref{tab:Step_shear_all} shows results for both geometries as well as for the PP geometry, as discussed in section 3.5 of the main text. The evolution of torque $M$ with time $t$ is shown in Figure~\ref{fig:Step_shear_vane_helix}. The data shows that the effect of shear history is much higher in the PP geometry as compared to vane and helix.
This difference is attributed to the local in-homogeneities in the paste (both in flow and composition) caused by the PP geometry, and that are discussed in detail in the main text.

\begin{table}[!h]
\centering
\begin{tabular}{ |c|c|c|c|c| } 
\hline
 \multirow{3}{4em}{Geometry} & \multirow{3}{4em}{Shear-1: Angular velocity $\Omega$ (rad/s)}   & \multirow{3}{4em}{Shear-2: Angular velocity $\Omega$ (rad/s)}  &
\multirow{3}{6em}{M (at the end of experiment) ~~(mN.m)} 
 \\
  & & &  \\
 & & &   \\
  & & &   \\
 & & &   \\
\hline
 PP & 3 & - & 1.314  \\ 
 PP & 0.6 & 3  & 0.667 \\ 
 PP & 6 & 3  & 1.738 \\
 Vane & 5.24 & - & 0.831  \\
 Vane & 1.05 & 5.24 &  0.826 \\
 Vane & 10.47 & 5.24 &  0.761 \\
 Helix & 5.24 & - & 2.682  \\
 Helix & 1.05 & 5.24 &  2.816 \\
 Helix & 10.47 & 5.24 &  2.778 \\
\hline
\end{tabular}
\captionsetup{width=31pc}
\caption{Values of torque $M$ at the end of each experiment for cement paste  $\phi$ = 0.45 (water-to-cement ratio w/c = 0.40) sheared using PP, vane and helix geometries by imposing various angular velocities $\Omega$ to the geometry. In some cases a single shear (shear 1) is applied for 300\,s and in other cases two consecutive rates (shear-1 and shear-2, each of 300\,s, with a time gap of 60\,s between the two) are imposed. }
\label{tab:Step_shear_all}
\end{table}

\section*{Competing Interests}

The authors declare no competing interests.

\section*{Acknowledgments}
The work was supported by Österreichische Bautechnik Vereinigung (ÖBV) and Österreichische Forschungsförderungsgesellschaft (FFG) (Project number: 870962). The authors acknowledge S.~Manneville for fruitful discussions and TU Wien Bibliothek for financial support through its Open Access Funding Programme.

%Bibliography
 % \bibliographystyle{ksfh_nat}  
 \bibliography{Dhar-sn-bibliography}

\newpage

\renewcommand{\thefigure}{S\arabic{figure}}
\setcounter{figure}{0}  

\section*{Supplementary Information - Discrepancies in Dynamic Yield Stress Measurements of Cement Pastes}

\maketitle

\section*{S1: Supplementary movie description: visualization of gap in parallel-plate geometry}
\label{supplement_video}
The movie shows the variation in the width of the moving band within the parallel plate geometry. The test protocol consists in imposing a stress ramp down, with the imposed stress value logarithmically decreasing with time  (see section 2.2.1 of main text). The cement paste of $\phi$ = 0.45 (w/c = 0.40) was used for the test.
The original movie was recorded at 10~fps with a resolution of 848x700~pixels and a spatial resolution of $\approx$ 230~pixel$/$mm.  The present video has been accelerated nine times the original speed and has been edited to include the real-time graph of varying stress and shear rate. The left side of the movie shows the real-time evolution of shear stress and shear rate as measured by the rheometer (calculated at $\frac{2}{3}$ of the radius). The video shows that, at high shear rate, the sample is apparently sheared within the whole gap. As the imposed shear stress decreases, the width of the moving band gradually decreases to reach a very small width near 1~$s^{-1}$, i.e., close to the defined yield stress. 
%The present video has been accelerated nine times the original speed and has been edited to include the real-time graph of varying stress and shear rate. 

\section*{S2: Schematic of measuring devices and geometries}
\label{schematic_geometries}

\subsection*{S2.1: Vane geometry}

\begin{figure*}[h!]
\centering
\includegraphics[width=29pc]{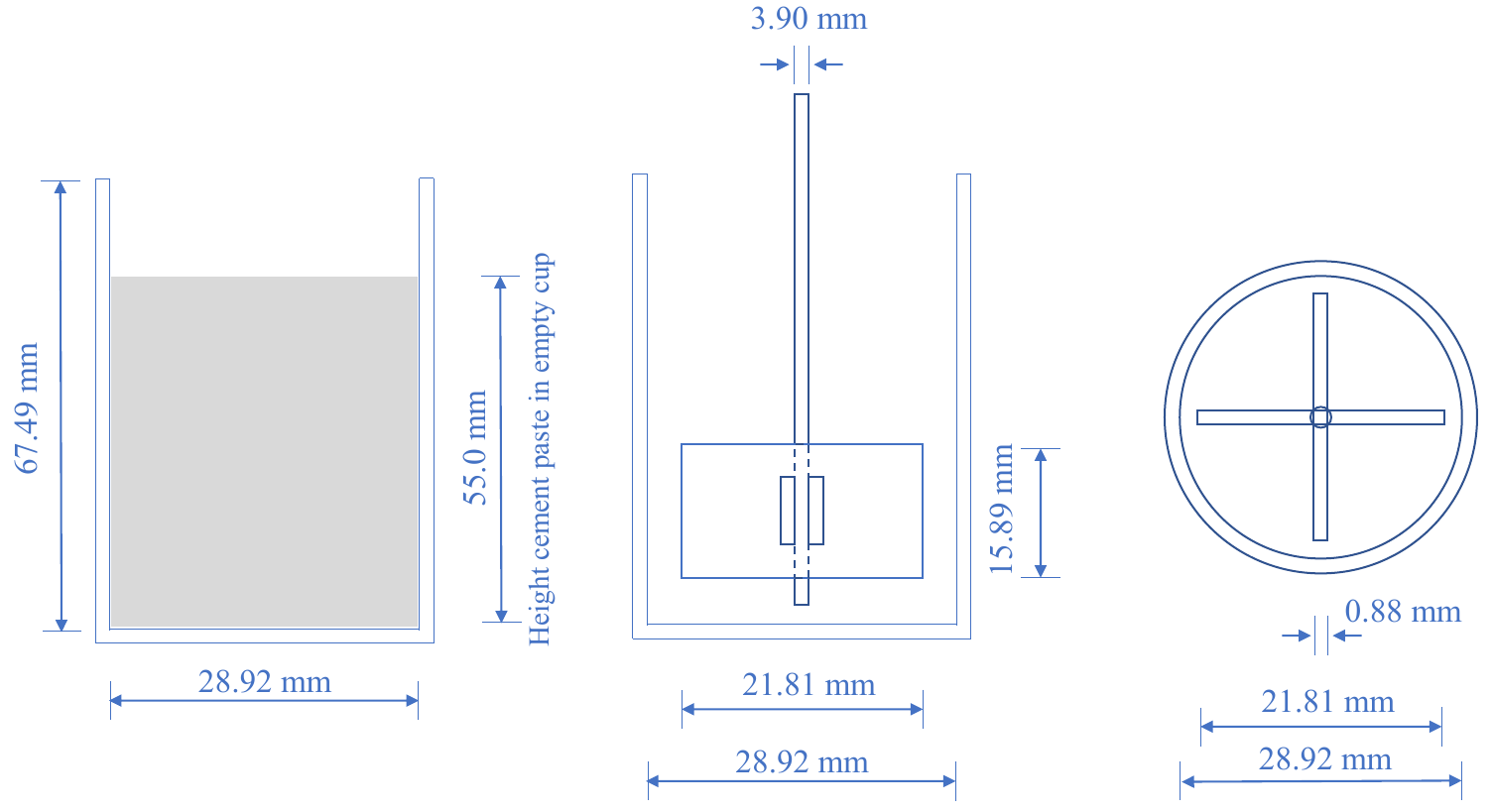}
\captionsetup{width=31pc}
\caption{Dimensions of vane geometry}
\label{fig:Vane_dimensions_Appendix}
\end{figure*}

The dimensions are shown in Fig. \ref{fig:Vane_dimensions_Appendix}. The surface of the outer cup was sandblasted.

\subsection*{S2.2: Helix geometry}

\begin{figure*}[t!]
\centering
\includegraphics[width=22pc]{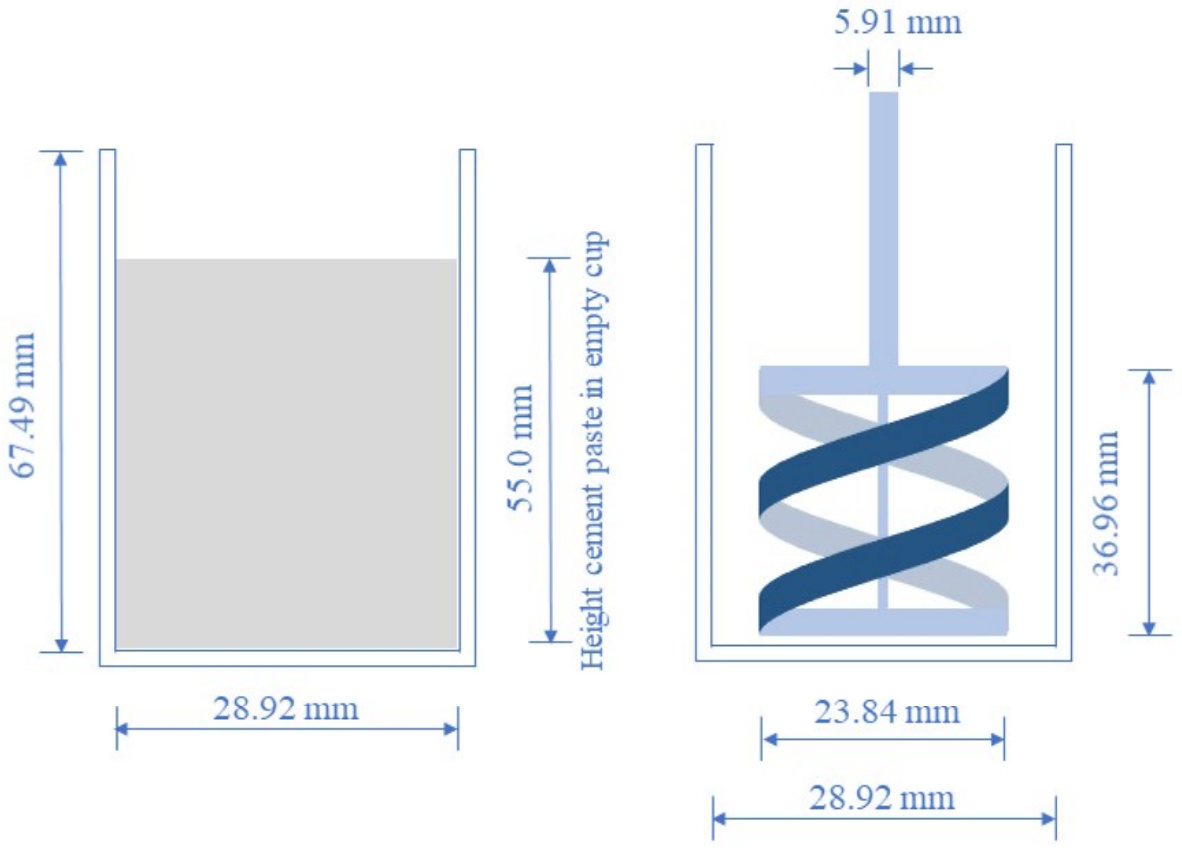}
\captionsetup{width=31pc}
\caption{Dimensions of helix geometry }
\label{fig:Helix_dimensions_Appendix}
\end{figure*}

The dimensions are shown in Fig. \ref{fig:Helix_dimensions_Appendix}. The surface of the outer cup was sandblasted.

\subsection*{S2.3: Serrated parallel plate geometry (PP)}

\begin{figure*}[t!]
\centering
\includegraphics[width=24pc]{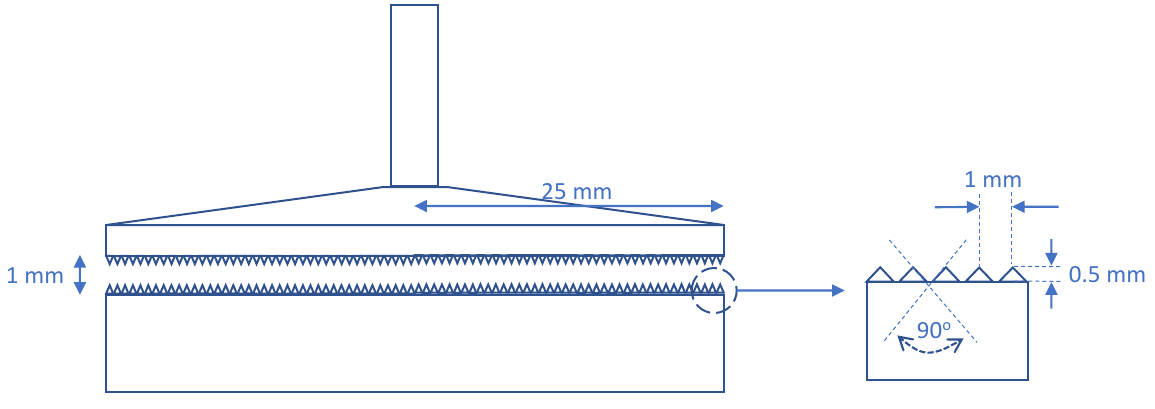}
\captionsetup{width=31pc}
\caption{Dimensions of parallel plate geometry (PP)}
\label{fig:PP50P2_dimensions_Appendix}
\end{figure*}

The dimensions are shown in Figure \ref{fig:PP50P2_dimensions_Appendix}. The bottom plate was custom-made for the study and matched the top plate in terms of its design. The equation that converts torque to stress is, for a Newtonian fluid:
\[ \tau_{max} = \frac{2M}{\pi R_{max}^3} \] where $R_{max}$ is the radius of the top moving plate, $\tau_{max}$ is the maximum stress at this radius and M is the torque \citep{macosko1994rheology}.

\subsection*{S2.4: Co-axial cylinder geometry}

\begin{figure*}[h!]
\centering
\includegraphics[width=22pc]{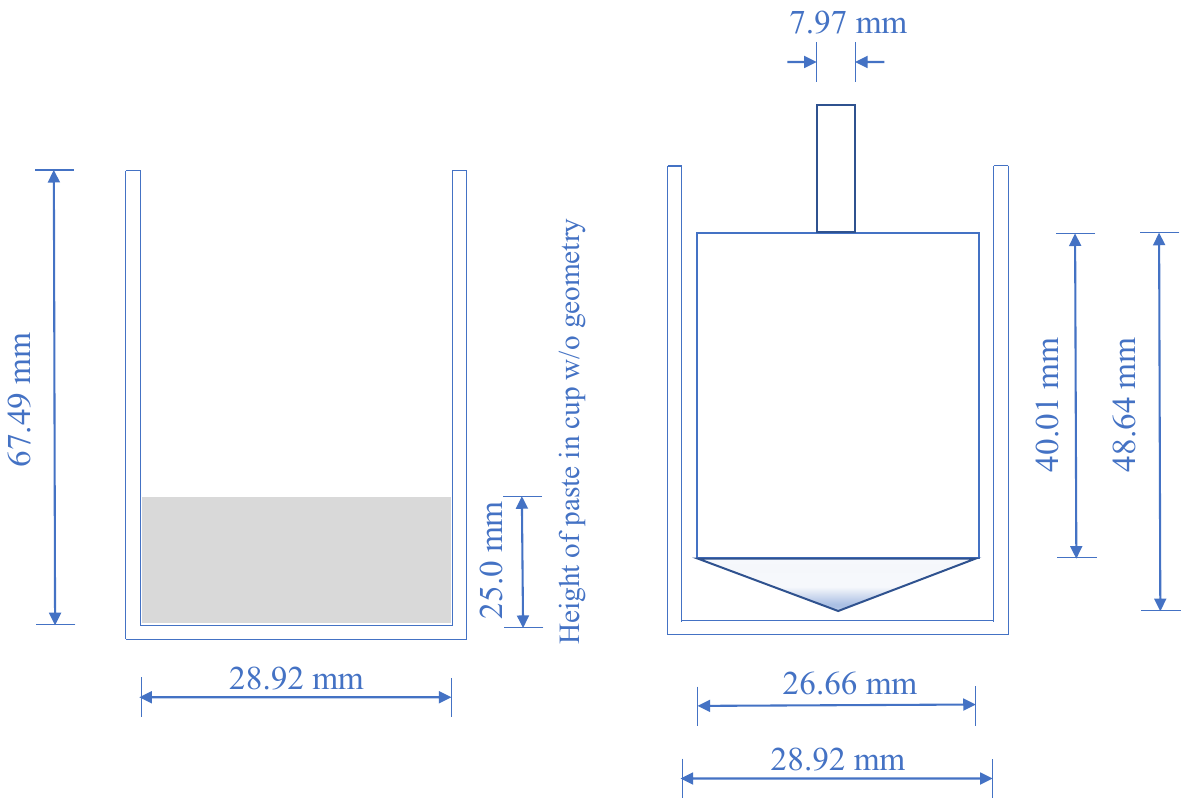}
\captionsetup{width=31pc}
\caption{Dimensions of co-axial cylinder geometry (CC) }
\label{fig:CC27S_dimensions_Appendix}
\end{figure*}

The dimensions are shown in Fig. \ref{fig:CC27S_dimensions_Appendix}. The surfaces of the rotating cylinder and the outer cup were sandblasted. 
The equation that converts torque to stress, is, for a Newtonian fluid: 
\[ \tau = K \cdot \frac{M}{ 2 \pi L R_i^2} \]   where 
$\tau$ is the stress, M the torque, L the height of the cylinder, $R_i$ the radius of the rotating cylinder and K is a correcting factor confidential to the manufacturer \citep{macosko1994rheology}. 

\subsection*{S2.5: Mini-cone for spread test}
\begin{figure*}[h!]
\centering
\includegraphics[width=8pc]{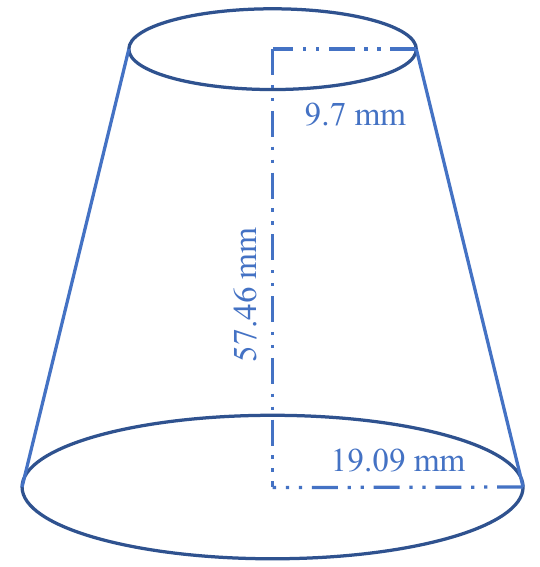}
\captionsetup{width=31pc}
\caption{Dimensions of mini-cone}
\label{fig:Mini_cone_dimensions_Appendix}
\end{figure*}

The dimensions are shown in Fig. \ref{fig:Mini_cone_dimensions_Appendix}.

\end{document}